\documentclass[twocolumn]{aastex701}

\usepackage{amsmath}
\usepackage{enumitem}
\setlist{nosep}
\usepackage[ruled,vlined,linesnumbered]{algorithm2e}

\SetCommentSty{smallcommentstyle}

\usepackage{tikz}
\usetikzlibrary{
  arrows.meta,
  positioning,
  shapes.geometric,
  fit,
  calc,
  decorations.pathreplacing,
  backgrounds,
  matrix
}
\usepackage{xcolor}

\definecolor{rank0}{RGB}{66,133,244}
\definecolor{rank1}{RGB}{234,67,53}
\definecolor{rank2}{RGB}{52,168,83}
\definecolor{avcolor}{RGB}{144,202,249}
\definecolor{thcolor}{RGB}{255,183,77}
\definecolor{partcolor}{RGB}{200,230,201}
\definecolor{mutexcolor}{RGB}{206,147,216}
\definecolor{rdmacolor}{RGB}{255,112,67}
\definecolor{arrowblue}{RGB}{30,90,180}
\definecolor{darkbg}{RGB}{248,248,255}
\definecolor{selfbg}{RGB}{230,240,255}
\definecolor{remotebg}{RGB}{255,245,230}

\newcommand{\drawHandler}[7]{%
  \begin{scope}[shift={(#1,#2)}]
    \fill[#3, rounded corners=3pt] (-2.3,-0.55) rectangle (2.3,1.25);
    \draw[gray!60, rounded corners=3pt] (-2.3,-0.55) rectangle (2.3,1.25);
    \node[font=\sffamily\tiny\bfseries, anchor=east] at (-1.55, 0.85) {P};
    \foreach \i in {0,...,7} {
      \pgfmathsetmacro{\xc}{-1.35 + \i*0.38}
      \pgfmathtruncatemacro{\isOcc}{(\i >= #6) * (\i < #7)}
      \ifnum \isOcc=1
        \fill[partcolor!80, draw=partcolor!30!black, semithick]
          (\xc-0.16, 0.69) rectangle (\xc+0.16, 1.01);
      \else
        \fill[partcolor!15, draw=partcolor!50!black, thin]
          (\xc-0.16, 0.69) rectangle (\xc+0.16, 1.01);
      \fi
    }
    \node[font=\sffamily\tiny\bfseries, arrowblue, anchor=east] at (-0.1, 0.35)
      {H=#6};
    \node[font=\sffamily\tiny\bfseries, rank1, anchor=west] at (0.1, 0.35)
      {T=#7};
    \node[font=\sffamily\tiny, gray!50, inner sep=2pt] at (0, -0.15)
      {SPSC (lock-free)};
  \end{scope}
}

\shorttitle{\textsc{STORM}: RDMA-based MC Transport for Distributed-memory Particle Simulations}
\shortauthors{Mizrachi, Raveh, \& Steinberg}

\begin{document}

\title{\textsc{STORM}: RDMA-based Monte Carlo Transport Scheme for Distributed-memory Particle Simulations}

\correspondingauthor{Maor Mizrachi}
\author{Maor Mizrachi}
\affiliation{School of Computer Science and Engineering, The Hebrew University, 9190401 Jerusalem, Israel}
\email[show]{maormiz@cs.huji.ac.il}

\author{Barak Raveh}
\affiliation{School of Computer Science and Engineering, The Hebrew University, 9190401 Jerusalem, Israel}
\email{barak.raveh@cs.huji.ac.il}

\author[0000-0003-0053-0696]{Elad Steinberg}
\affiliation{Racah Institute of Physics, The Hebrew University, 9190401 Jerusalem, Israel}
\email{elad.steinberg@mail.huji.ac.il}

\received{2026 July 20}
\revised{2026 September 1}
\accepted{2026 September 3}
\published{2026 September 29}

\begin{abstract}
Monte Carlo particle transport enables high-fidelity astrophysical radiation and neutrino simulations---from core-collapse supernovae and neutron-star mergers to accretion flows---by handling multidimensional geometries, frequency dependence, and moving media without angular discretization. However, inter-rank communication limits scalability on unstructured meshes: standard two-sided MPI requires receivers to post ``receives'' and ``poll'' completions, creating per-iteration progress overhead that grows with the number of communication partners. Such problems have not demonstrated high scaling efficiency at $O(10^{4})$ cores. We present \textsc{STORM} (Scalable Transport via One-sided Remote Memory), an open-source library for Monte Carlo transport on general meshes, physics, and boundary conditions. \textsc{STORM} provides a lock-free, mesh-independent communication layer that replaces MPI's matched-send/receive semantics with remote direct memory access (RDMA)---one-sided operations that write directly into a remote rank's memory without involving its CPU. Each rank pair shares a single-producer, single-consumer ring buffer; RDMA writes transfer particles while receivers remain passive. A two-sided MPI backend provides a portable fallback. In an adversarial uniform-emission benchmark, the RDMA backend sustains $>97\%$ weak-scaling and $>88\%$ work-preserving scaling efficiency up to 13,440~cores (112~cores per network adapter), with $1.14$--$1.27\times$ speedups over the two-sided alternative. In a Hohlraum \textsc{IMC} benchmark at 4480~ranks, it is $1.41\times$ faster because MPI progress overhead is reduced by $6.1\times$. By decoupling communication from physics models and mesh representations, \textsc{STORM} removes a barrier to scaling Monte Carlo transport in astrophysical multiphysics codes, enabling coupled radiation-hydrodynamics with energy- and angle-resolved photon or neutrino transport on dynamically evolving meshes at scale.
\end{abstract}

\keywords{Radiative transfer simulations (1967) --- Radiative transfer (1335) --- Computational methods (1965) --- Distributed computing (1971) --- Monte Carlo methods (2238)}


\section{Introduction}
Radiation and neutrino transport are central to many of the most challenging problems in computational astrophysics. In core-collapse supernovae, neutrino heating behind the stalled shock helps determine whether an explosion is launched; in neutron-star mergers and post-merger accretion flows, neutrino cooling and lepton-number transport shape the composition of the ejecta and the resulting electromagnetic counterparts; in supernova ejecta, stellar envelopes, and accretion disks, radiation transport controls thermal histories, spectra, and the exchange of momentum between matter and radiation \citep{Kasen2006,Noebauer2019,Mezzacappa2020,Foucart2023}. These systems are difficult to model because the radiation/neutrino field is intrinsically high dimensional, depending on space, direction, frequency or energy, and time, while also coupling nonlinearly to matter through opacities, emissivities, scattering kernels, and velocity-dependent effects.

Monte Carlo transport is attractive in this setting because it samples particle histories directly rather than explicitly discretizing the full angular phase space. This makes it well suited to multidimensional geometries, anisotropic radiation fields, frequency dependence, and moving media. Astrophysical Monte Carlo radiative-transfer codes such as \textsc{SEDONA}, \textsc{TORUS}, \textsc{MCFOST}, and \textsc{ARTEMIS} have demonstrated the method's flexibility for spectra, continuum and line transfer, dust radiative equilibrium, polarization, and time-dependent transport \citep{Kasen2006,Pinte2006,Harries2019,Noebauer2019,Ryan2024}. For radiation hydrodynamics, Implicit Monte Carlo \citep[\textsc{IMC};][]{FleckCummings1971,Wollaber2016} and Discrete Diffusion Monte Carlo \citep[\textsc{DDMC};][]{Densmore2007,Densmore2012} provide stable and efficient treatments of optically thick thermal radiation, and recent moving-mesh work such as \textsc{AREPO-MCRT} \citep{Smith2020} illustrates the growing role of Monte Carlo transport in dynamic, unstructured astrophysical simulations.

The same considerations motivate Monte Carlo methods for neutrino transport. Fully resolved Boltzmann or discrete-ordinates neutrino transport remains extremely expensive in multidimensional supernova and merger calculations, while leakage, flux-limited diffusion, and moment schemes trade angular fidelity for computational speed \citep{Messer1998,Richers2017,Mezzacappa2020,Foucart2023}. Moment methods with analytic closures, including \textsc{M1}, are now widely used, but their closures can fail when the radiation field contains intersecting beams or strongly nonlocal angular structure \citep{Murchikova2017,Foucart2018}. Monte Carlo neutrino transport has, therefore, become an important complementary approach for benchmark calculations, snapshot transport, closure construction, and problems where angle- and energy-dependent neutrino physics is especially important \citep{Abdikamalov2012,Richers2015,Foucart2018,Foucart2023}.

As Monte Carlo transport is coupled to larger three-dimensional astrophysical simulations, the communication problem becomes a limiting concern. In a domain-decomposed calculation, photon or neutrino packets are advanced locally until they cross a subdomain boundary, at which point ownership must be transferred to another MPI rank. The resulting traffic is irregular: packets cross boundaries at stochastic times, in directions set by the radiation field and mesh geometry, and with rates that can vary strongly across the domain. Moving meshes, adaptive refinement, and load rebalancing make the pattern even less predictable. If communication requires the receiving rank to poll all neighbors or to participate in matched receives, particle transfer can compete directly with useful transport work and can become a scalability bottleneck even when the underlying physical Monte Carlo algorithm is highly parallel.

One way to reduce this overhead is to bypass the matched-message semantics of MPI and instead use one-sided remote direct memory access (RDMA) operations, which allow a sender to deposit data directly into a remote rank's memory without involving the remote CPU. The relevant distinction, however, is not simply ``RDMA versus MPI.'' Optimized MPI point-to-point implementations commonly use RDMA-capable transports underneath, for example, through rendezvous protocols or UCX. The overhead addressed here comes from the abstraction boundary: MPI exposes a generic matched-message interface in which receives must be posted, completion and progress are driven through tests, waits, or probes, and memory-registration and ordering details are hidden behind a portable library. For stochastic particle transport, this generic asynchronous-progress model can put probing in the hot loop and gives the application little control over the fine-grained details that determine latency under bursty, neighbor-dependent traffic. Our approach is, therefore, to replace the particle communication library with an application-specific low-level RDMA protocol, while retaining an optimized MPI point-to-point backend as the portable baseline.

Previous work has addressed related communication issues in large-scale Monte Carlo transport. \citet{Brunner2009} introduced an efficient domain-decomposition algorithm with a tree-based completion detector for particle Monte Carlo, while \citet{Romano2011} explored MPI remote memory access for particle communication in Monte Carlo neutron transport, and \citet{Romano2013} studied tally server decomposition. Much of the transport literature, however, necessarily emphasizes the physics model, acceleration strategy, or closure approximation. The communication substrate itself remains a reusable systems problem: ranks must move particles asynchronously, coordinate remote buffers safely, avoid deadlock, resize buffers under bursty traffic, and detect global completion without excessive synchronization.

In this paper, we present \textsc{STORM}\footnote{\url{https://github.com/maormizrachi/STORM}} (scalable transport via one-sided remote memory), a publicly available, open-source library that provides a mesh-independent communication and particle-management layer for distributed Monte Carlo transport. \textsc{STORM} implements Monte Carlo transport of general physics on general meshes with general boundary conditions, using direct one-sided RDMA operations, pairwise communication Handlers, local aggregation of outgoing particles, a deadlock-free reallocation protocol, and a distributed termination detector. Its only physics-side requirement is that the application determine when a packet leaves the local subdomain and identify the destination rank and associated metadata. \textsc{STORM} is distributed as a standalone package and is also integrated into the open-source three-dimensional moving-mesh hydrodynamics code \textsc{RICH}\footnote{\url{https://gitlab.com/eladtan/RICH}} \citep{Yalinewich2015}, where it connects to the \textsc{MadVoro} parallel Voronoi tessellation builder \citep{Mizrachi2025madvoro} for Monte Carlo radiative transfer on moving Voronoi meshes. The \textsc{IMC} implementation built on \textsc{STORM} is described separately in a companion paper (M.~Mizrachi et al. 2026, in preparation). Here we deliberately isolate the communication layer, which is applicable to photon, neutrino, and other particle-transport problems on structured, adaptive, unstructured, or moving-mesh decompositions.

The remainder of this paper is organized as follows. Section~\ref{sec:mpi_background} describes the Open Fabrics Interfaces (OFI) application programming interface (API) used by our primary backend and briefly introduces the alternative backends. Section~\ref{sec:method} presents \textsc{STORM}'s communication architecture in detail: the per-pair Handler design, the one-sided transfer protocol, particle aggregation, deadlock-free buffer reallocation, and distributed termination detection. The coupling to a moving Voronoi mesh, including remapping and load balancing, is described in Section~\ref{sec:moving_mesh}. Section~\ref{sec:benchmarking} validates the scheme on a uniform-emission benchmark, presents constant-workload-scaling and weak-scaling results, and introduces a cylindrical Hohlraum benchmark for realistic \textsc{IMC} timing comparisons. Section~\ref{sec:conclusion} concludes with a summary and directions for future work.

\section{RDMA and Communication Backends}\label{sec:mpi_background}
RDMA allows one process to read from or write to the memory of a remote process without involving the remote CPU. On modern High Performance Computing (HPC) clusters, the network adapter executes RDMA operations entirely in hardware after the application posts a request, delivering low-latency, high-throughput data movement that is ideally suited for the fine-grained, asynchronous particle transfers required by distributed Monte Carlo transport.

RDMA is not an exotic technology external to the MPI ecosystem. On the contrary, most modern MPI implementations already use RDMA internally---for example, via rendezvous protocols for large messages or through the UCX framework---and MPI~3 exposes one-sided RDMA semantics directly through its RMA interface (\texttt{MPI\_Put}, \texttt{MPI\_Get}, \texttt{MPI\_Compare\_and\_swap}). The relevant distinction in this work is, therefore, not RDMA-capable hardware versus non-RDMA hardware, but an application-specific low-level RDMA protocol versus generic MPI communication semantics. Our OFI backend bypasses the MPI abstraction layer to issue RDMA operations directly through the Open Fabrics Interfaces (\texttt{libfabric}) API, gaining fine-grained control over memory registration, completion detection, operation ordering, and progress that the standardized MPI interface cannot fully expose. This section describes the OFI interface, followed by a brief overview of the alternative backends.

\subsection{Open Fabrics Interfaces (OFI)}\label{sec:ofi_overview}
\begin{figure*}
    \centering
    \resizebox{\textwidth}{!}{\begin{tikzpicture}[>=Stealth, font=\sffamily\small,
    box/.style={draw, thick, rounded corners=3pt, minimum height=0.7cm,
                inner sep=4pt, font=\sffamily\scriptsize},
    mrbox/.style={box, fill=partcolor!40, draw=partcolor!70!black,
                  minimum width=3.2cm},
    epbox/.style={box, fill=avcolor!50, draw=avcolor!70!black,
                  minimum width=3.2cm},
    avbox/.style={box, fill=avcolor!20, draw=avcolor!70!black,
                  minimum width=3.2cm},
    cqbox/.style={box, fill=thcolor!40, draw=thcolor!70!black,
                  minimum width=3.2cm},
    dombox/.style={draw=rank0!60, thick, rounded corners=5pt,
                  fill=rank0!4, inner sep=6pt},
    rankbox/.style={draw=gray!60, thick, rounded corners=6pt,
                    fill=darkbg, inner sep=8pt},
    nicbox/.style={box, fill=rdmacolor!25, draw=rdmacolor!70!black,
                   minimum width=3.6cm, minimum height=0.6cm},
    netlabel/.style={font=\sffamily\small\bfseries, gray!60},
    stepcirc/.style={shape=circle, draw=rank2!80!black, thick, inner sep=0pt,
                     minimum size=14pt, font=\sffamily\tiny\bfseries,
                     fill=white},
  ]

  \begin{scope}[shift={(1,0)}]
    \node[rankbox, minimum width=6.0cm, minimum height=7.5cm] (rankA) at (0, 3.5) {};
    \node[font=\sffamily\bfseries, rank0] at (0, 8.65) {Rank A (origin)};

    \node[box, fill=selfbg, draw=rank0!50, minimum width=3.6cm]
      (cpuA) at (0, 7.8) {Application (user space)};

    \node[dombox, minimum width=5.2cm, minimum height=5.6cm]
      (domA) at (0, 4.1) {};
    \node[font=\sffamily\scriptsize\bfseries, rank0!80!black, anchor=north west]
      at ([xshift=3pt, yshift=-2pt] domA.north west) {Domain};

    \node[mrbox] (mr1A) at (0, 6.0) {Buffer A \quad\textsf{\tiny mr\_key}};
    \node[mrbox] (mr2A) at (0, 5.15) {Buffer B \quad\textsf{\tiny mr\_key}};
    \node[font=\sffamily\tiny\bfseries, partcolor!60!black, anchor=east]
      at ([xshift=-4pt] mr1A.west) {MR};
    \node[font=\sffamily\tiny\bfseries, partcolor!60!black, anchor=east]
      at ([xshift=-4pt] mr2A.west) {MR};

    \node[epbox] (epA) at (0, 4.1) {Endpoint \textsf{\tiny(RDM)}};
    \node[avbox] (avA) at (0, 3.25) {Address Vector \textsf{\tiny[B, C, \ldots]}};
    \node[font=\sffamily\tiny\bfseries, avcolor!60!black, anchor=east]
      at ([xshift=-4pt] epA.west) {EP};
    \node[font=\sffamily\tiny\bfseries, avcolor!60!black, anchor=east]
      at ([xshift=-4pt] avA.west) {AV};

    \node[cqbox] (cqA) at (0, 2.15) {Completion Queue};
    \node[font=\sffamily\tiny\bfseries, thcolor!60!black, anchor=east]
      at ([xshift=-4pt] cqA.west) {CQ};

    \node[nicbox] (nicA) at (0, 0.5) {NIC (fabric provider)};
  \end{scope}

  \begin{scope}[shift={(10.5,0)}]
    \node[rankbox, minimum width=6.0cm, minimum height=7.5cm] (rankB) at (0, 3.5) {};
    \node[font=\sffamily\bfseries, rank1] at (0, 8.65) {Rank B (target)};

    \node[box, fill=gray!10, draw=gray!40, minimum width=3.6cm, text=gray!50]
      (cpuB) at (0, 7.8) {CPU uninvolved};

    \node[dombox, minimum width=5.2cm, minimum height=5.6cm]
      (domB) at (0, 4.1) {};
    \node[font=\sffamily\scriptsize\bfseries, rank0!80!black, anchor=north west]
      at ([xshift=3pt, yshift=-2pt] domB.north west) {Domain};

    \node[mrbox, fill=partcolor!60] (mr1B) at (0, 6.0)
      {Buffer X \quad\textsf{\tiny mr\_key}};
    \node[mrbox] (mr2B) at (0, 5.15) {Buffer Y \quad\textsf{\tiny mr\_key}};
    \node[font=\sffamily\tiny\bfseries, partcolor!60!black, anchor=east]
      at ([xshift=-4pt] mr1B.west) {MR};
    \node[font=\sffamily\tiny\bfseries, partcolor!60!black, anchor=east]
      at ([xshift=-4pt] mr2B.west) {MR};

    \node[epbox, fill=avcolor!20] (epB) at (0, 4.1) {Endpoint \textsf{\tiny(RDM)}};
    \node[avbox] (avB) at (0, 3.25) {Address Vector \textsf{\tiny[A, C, \ldots]}};
    \node[font=\sffamily\tiny\bfseries, avcolor!60!black, anchor=east]
      at ([xshift=-4pt] epB.west) {EP};
    \node[font=\sffamily\tiny\bfseries, avcolor!60!black, anchor=east]
      at ([xshift=-4pt] avB.west) {AV};

    \node[cqbox, fill=thcolor!15] (cqB) at (0, 2.15) {Completion Queue};
    \node[font=\sffamily\tiny\bfseries, thcolor!60!black, anchor=east]
      at ([xshift=-4pt] cqB.west) {CQ};

    \node[nicbox] (nicB) at (0, 0.5) {NIC (fabric provider)};
  \end{scope}

  \draw[gray!30, very thick, dashed] (5.75, -0.5) -- (5.75, 8.8);
  \node[netlabel] at (5.75, 9.0) {Network fabric};


  \draw[->, very thick, rank2!80!black, densely dashed]
    (cpuA.east) .. controls (5.0, 7.0) and (5.0, 4.8) .. (epA.east)
    node[stepcirc, pos=0.4] (s1) {1};
  \node[font=\sffamily\tiny, rank2!70!black, anchor=west]
    at ([xshift=3pt] s1.east) {\texttt{fi\_write}};

  \draw[->, very thick, rank2!80!black]
    (mr1A.south west) .. controls (-3.5, 6.0) and (-3.5, 0.5) .. (nicA.north west)
    node[stepcirc, pos=0.5] (s2) {2};
    \node[font=\sffamily\tiny, rank2!70!black, anchor=east]
    at ([xshift=-3pt, yshift=4pt] s2.west) {NIC reads};
  \node[font=\sffamily\tiny, rank2!70!black, anchor=east]
    at ([xshift=-3pt, yshift=-5pt] s2.west) {local MR};

  \draw[->, very thick, rank2!80!black]
    (nicA.east) -- (nicB.west)
    node[stepcirc, pos=0.35] (s3) {3};
  \node[font=\sffamily\tiny, rank2!70!black, anchor=south]
    at ([yshift=3pt] s3.north) {data over fabric};
  \draw[->, very thick, rank2!80!black]
    (nicB.north west) .. controls (7.0, 1.0) and (7.0, 5.5) .. (mr1B.south west);

  \draw[->, thick, thcolor!80!black, dotted]
    (nicA.north) .. controls (2.8, 1.1) and (2.8, 1.6) .. (cqA.south)
    node[stepcirc, pos=0.5] (s4) {4};
  \node[font=\sffamily\tiny, thcolor!70!black, anchor=west]
    at ([xshift=3pt, yshift=-3pt] s4.east) {\texttt{fi\_cq\_read}};

  \begin{scope}[shift={(0.3, -2.2)}]
    \node[font=\sffamily\scriptsize\bfseries, anchor=west] at (0, 0) {Legend:};
    \draw[->, thick, rank2!80!black, densely dashed] (1.6, 0) -- (2.4, 0);
    \node[font=\sffamily\scriptsize, anchor=west] at (2.5, 0) {CPU $\to$ EP (\texttt{fi\_write})};
    \draw[->, thick, rank2!80!black] (6.2, 0) -- (7.0, 0);
    \node[font=\sffamily\scriptsize, anchor=west] at (7.1, 0) {NIC data path};
    \draw[->, thick, thcolor!80!black, dotted] (9.6, 0) -- (10.4, 0);
    \node[font=\sffamily\scriptsize, anchor=west] at (10.5, 0) {completion};
  \end{scope}

\end{tikzpicture}}
    \caption{OFI (\texttt{libfabric}) architecture and one-sided RDMA write in four steps. (1)~The application on Rank~A issues an RDMA operation (e.g., \texttt{fi\_write}), specifying the local buffer's memory descriptor, the remote buffer's address and memory-region key, and the target's \texttt{fi\_addr\_t} from the Address Vector. (2)~The NIC reads the source data from the local registered Memory Region. (3)~The NIC transmits the data over the network fabric and writes it directly into Rank~B's registered Memory Region---without involving Rank~B's CPU. (4)~A completion entry is posted to Rank~A's Completion Queue, which the application polls via \texttt{fi\_cq\_read} to detect that the operation has finished. Each rank maintains a single reliable datagram (RDM) endpoint shared across all peers; remote ranks are addressed through an address vector. Memory regions and the endpoint are grouped within a domain for resource scoping and access control.}
    \label{fig:ofi_architecture}
\end{figure*}
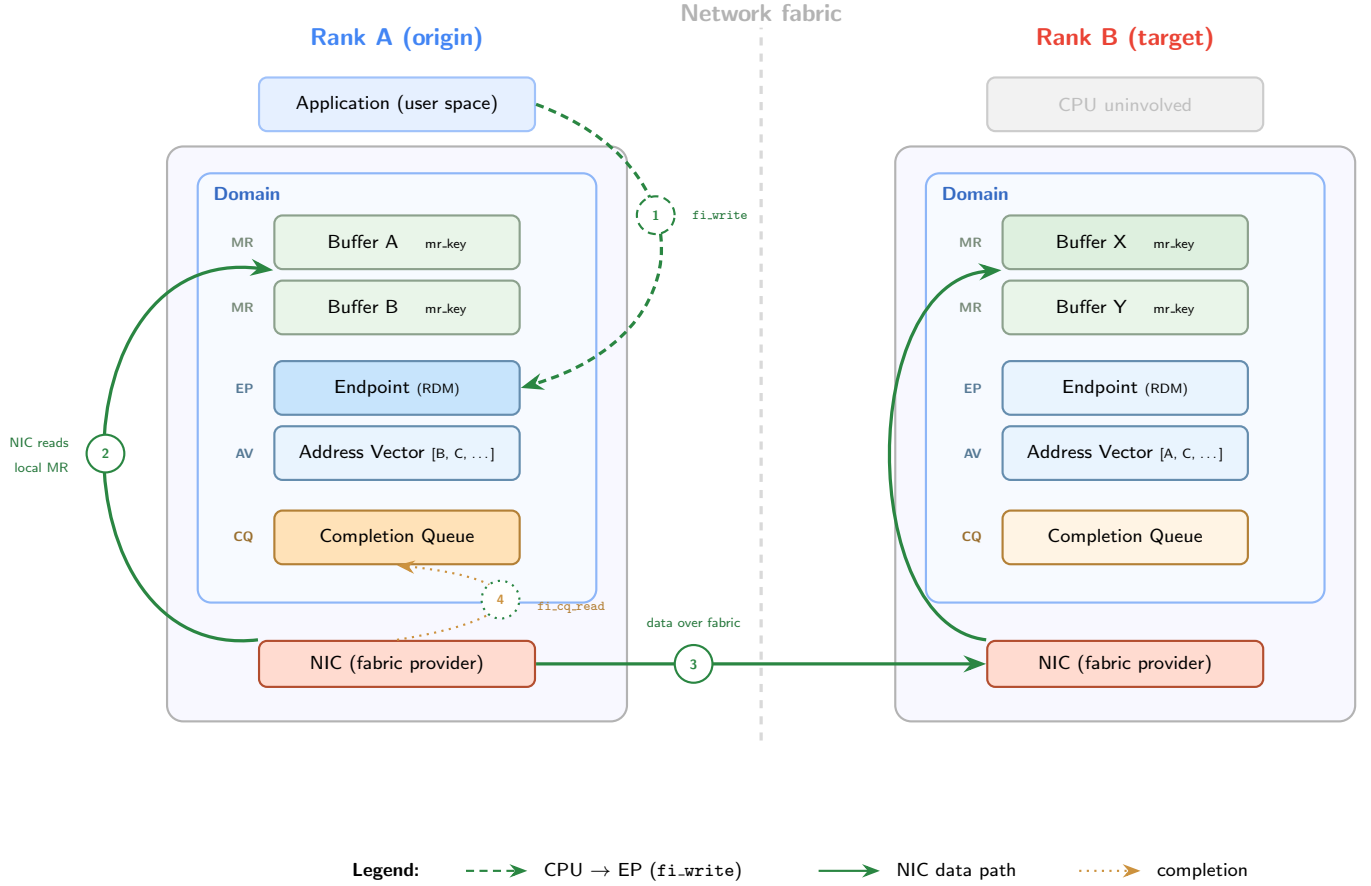

The Open Fabrics Interfaces API (\texttt{libfabric}) provides a portable, provider-agnostic interface for issuing RDMA operations across diverse network fabrics. Unlike InfiniBand Verbs, which exposes hardware-specific abstractions tied to InfiniBand adapters, \texttt{libfabric} supports multiple fabric providers---including InfiniBand (via the \texttt{verbs} provider), Cray Slingshot (\texttt{cxi}), AWS Elastic Fabric Adapter (\texttt{efa}), and Ethernet (\texttt{tcp/rxm})---through a unified API. At startup, \textsc{STORM} queries the available providers and automatically selects the best match for the local hardware, requiring no recompilation or user intervention. Figure~\ref{fig:ofi_architecture} summarizes the key abstractions.

The resource hierarchy begins with a \emph{fabric} object representing the physical network, from which a \emph{domain} is created to scope all communication resources. Memory targeted by RDMA must be \emph{registered} with the domain via \texttt{fi\_mr\_reg}, which pins the pages in physical memory and returns a memory-region key. Each Handler registers two regions: the particle ring buffer and the head/tail counter pair.

Communication proceeds through a single \emph{reliable datagram} (RDM) endpoint shared across all peers. Unlike connected endpoint models that require one endpoint per peer, the RDM endpoint is connectionless: remote ranks are addressed through an \emph{address vector}---a lookup table of peer addresses populated once at setup time via \texttt{fi\_av\_insert}. This design makes the endpoint count independent of the number of communication partners, improving scalability for many-neighbor decompositions. Because all of our operations are one-sided, the remote endpoint need not execute any receive call. A single shared completion queue, polled via \texttt{fi\_cq\_read}, detects operation completions.

Our implementation uses four RDMA operation types, issued as direct function calls on the endpoint: \emph{RDMA Write} (\texttt{fi\_write}) to transfer particle data into the remote ring buffer, \emph{RDMA Read} (\texttt{fi\_read}) to read remote head/tail counters, \emph{Atomic Fetch-and-Add} (\texttt{fi\_fetch\_atomic}) to publish newly written particles (Section~\ref{sec:transfer}), and \emph{Atomic Compare-and-Swap} (\texttt{fi\_compare\_atomic}) for the reallocation protocol (Section~\ref{sec:reallocation}). At startup, each rank obtains its local endpoint name via \texttt{fi\_getname}; the names are exchanged via \texttt{MPI\_Allgather} and inserted into the address vector. This one-time setup requires no per-peer state-machine transitions and is amortized over subsequent transfers.

\subsection{Alternative Backends}\label{sec:intro_backends}
Our implementation supports three additional communication backends beyond OFI: a native InfiniBand Verbs (IBV) backend, MPI's standard RMA interface, and classical two-sided (point-to-point) MPI messaging \citep{Brunner2009}. All four backends implement the same internal interface, so the choice is transparent to application code and can be changed at build or runtime.

The \emph{IBV backend} issues RDMA operations directly through the InfiniBand Verbs API (\texttt{libibverbs}), using one reliable connected (RC) queue pair per peer rank. It provides the same fine-grained control as OFI but is specific to InfiniBand hardware; on InfiniBand clusters, it offers comparable performance to the OFI backend with the \texttt{verbs} provider.

The \emph{MPI RMA backend} uses the standard one-sided primitives (\texttt{MPI\_Put}, \texttt{MPI\_Get}, \texttt{MPI\_Compare\_and\_swap}) within passive-target epochs (\texttt{MPI\_Win\_lock}/\texttt{MPI\_Win\_unlock}), providing a portable fallback that runs on any MPI-3-compliant implementation without requiring direct access to the network hardware. In practice, however, the MPI RMA API imposes several restrictions that limit its effectiveness for fine-grained particle transfers: operations must be bracketed by epochs, completion is coarse-grained (an entire epoch rather than individual operations), and the interplay between memory models, synchronization modes, and progress semantics constrains the ability of MPI libraries to map RMA calls efficiently onto the underlying hardware. As a result, the MPI RMA backend is consistently the slowest of our backends and is included primarily as a lowest-common-denominator fallback.

The \emph{two-sided (point-to-point) backend} replaces all one-sided operations with matched \texttt{MPI\_Isend}/\texttt{MPI\_Irecv} pairs. Although conceptually simpler, this backend has been carefully optimized: it uses nonblocking operations with manual progress, message aggregation, and per-peer send queues to minimize synchronization overhead. It serves as a universal fallback that requires no special hardware or MPI~RMA support. This is a conservative baseline: the MPI library may still use RDMA internally, but the application must drive progress through the point-to-point interface and cannot directly control remote buffers, per-operation completion, or the ordering of low-level network operations. We present comparative results between the OFI and point-to-point backends in Section~\ref{sec:benchmarking}.

\section{Communication Architecture}\label{sec:method}
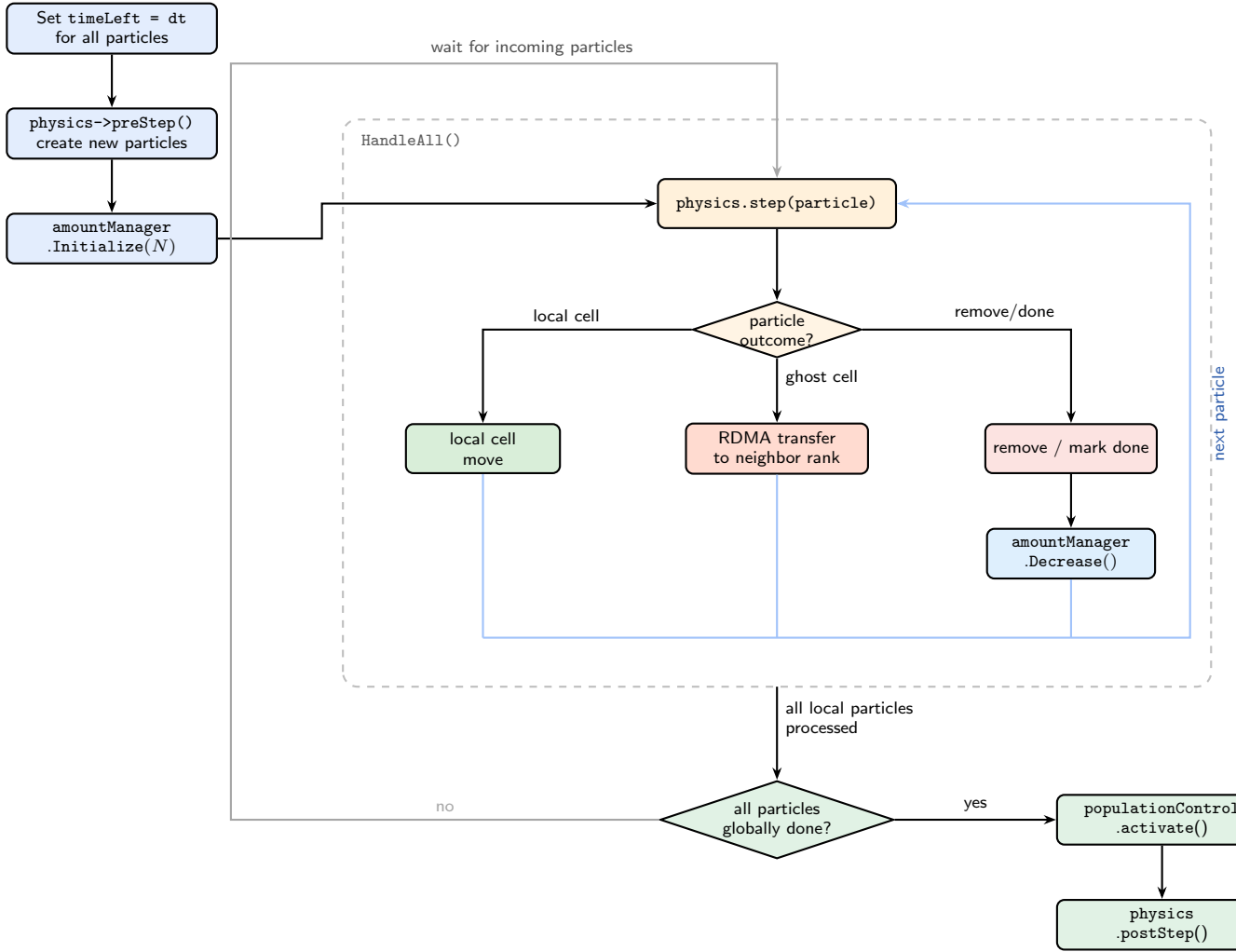
\begin{figure*}
    \centering
    \resizebox{\textwidth}{!}{\begin{tikzpicture}[>=Stealth, font=\sffamily\small]

  \begin{scope}[shift={(0, -12)}]
    \tikzset{
      fbox/.style={
        draw, rounded corners=3pt, thick, minimum width=3.0cm,
        minimum height=0.7cm, font=\sffamily\scriptsize, align=center
      },
      dec/.style={
        draw, diamond, thick, aspect=2.5, inner sep=1pt,
        font=\sffamily\scriptsize, align=center
      }
    }

    \node[fbox, fill=rank0!15] (init) at (-2.5, 10.0)
      {Set \texttt{timeLeft = dt}\\for all particles};
    \node[fbox, fill=rank0!15] (prestep) at (-2.5, 8.5)
      {\texttt{physics->preStep()}\\create new particles};
    \node[fbox, fill=rank0!15] (initcnt) at (-2.5, 7.0)
      {\texttt{amountManager}\\.\texttt{Initialize}$(N)$};

    \draw[-{Stealth[length=5pt]}, thick] (init) -- (prestep);
    \draw[-{Stealth[length=5pt]}, thick] (prestep) -- (initcnt);

    \def\boxL{0.8}
    \def\boxR{13.2}
    \def\boxT{8.7}
    \def\boxB{0.6}

    \draw[dashed, thick, gray!50, rounded corners=6pt]
      (\boxL, \boxB) rectangle (\boxR, \boxT);
    \node[font=\sffamily\scriptsize\bfseries, anchor=north west, gray!70!black]
      at (\boxL+0.15, \boxT-0.08) {\texttt{HandleAll()}};


    \node[fbox, fill=thcolor!20, minimum width=3.4cm] (step) at (7, 7.5)
      {\texttt{physics.step(particle)}};

    \draw[-{Stealth[length=5pt]}, thick]
      (initcnt.east) -- (\boxL-0.3, 7.0) |- (step.west);

    \node[dec, fill=thcolor!15, inner sep=0pt, aspect=3.0]
      (fate) at (7, 5.7) {particle\\outcome?};

    \draw[-{Stealth[length=5pt]}, thick] (step) -- (fate);


    \node[fbox, fill=partcolor!70, minimum width=2.2cm] (local) at (2.8, 4.0)
      {local cell\\move};
    \draw[-{Stealth[length=5pt]}, thick] (fate.west) -|
      node[above left, font=\sffamily\scriptsize, pos=0.2] {local cell} (local.north);

    \node[fbox, fill=rdmacolor!25, minimum width=2.6cm] (xfer) at (7, 4.0)
      {RDMA transfer\\to neighbor rank};
    \draw[-{Stealth[length=5pt]}, thick] (fate.south) --
      node[right, font=\sffamily\scriptsize, pos=0.3] {ghost cell} (xfer.north);

    \node[fbox, fill=rank1!15, minimum width=2.4cm] (rem) at (11.2, 4.0)
      {remove / mark done};
    \draw[-{Stealth[length=5pt]}, thick] (fate.east) -|
      node[above right, font=\sffamily\scriptsize, pos=0.2] {remove/done} (rem.north);

    \node[fbox, fill=avcolor!30, minimum width=2.4cm] (decr) at (11.2, 2.5)
      {\texttt{amountManager}\\.\texttt{Decrease}$()$};
    \draw[-{Stealth[length=5pt]}, thick] (rem) -- (decr);

    \def\innerBusY{1.3}
    \draw[thick, rank0!50] (local.south) -- (2.8, \innerBusY);
    \draw[thick, rank0!50] (xfer.south) -- (7, \innerBusY);
    \draw[thick, rank0!50] (decr.south) -- (11.2, \innerBusY);
    \draw[thick, rank0!50] (2.8, \innerBusY) -- (11.2, \innerBusY);
    \draw[-{Stealth[length=5pt]}, thick, rank0!50]
      (11.2, \innerBusY) -- (\boxR-0.3, \innerBusY) -- (\boxR-0.3, 7.5) -- (step.east);
    \node[font=\sffamily\scriptsize, rank0!70!black, anchor=west] at (\boxR-0.1, 4.5)
      {\rotatebox{90}{next particle}};

    \node[dec, fill=rank2!15, aspect=3.0] (gdone) at (7, -1.3)
      {all particles\\globally done?};

    \draw[-{Stealth[length=5pt]}, thick] (7, \boxB) --
      node[right, font=\sffamily\scriptsize, pos=0.35, align=left]
        {all local particles\\processed} (gdone.north);

    \node[fbox, fill=rank2!15] (popctl) at (12.5, -1.3)
      {\texttt{populationControl}\\.\texttt{activate}()};
    \node[fbox, fill=rank2!15] (post) at (12.5, -2.8)
      {\texttt{physics}\\.\texttt{postStep}()};

    \draw[-{Stealth[length=5pt]}, thick] (gdone.east) --
      node[above, font=\sffamily\scriptsize] {yes} (popctl.west);
    \draw[-{Stealth[length=5pt]}, thick] (popctl) -- (post);

    \def\outerLoopX{-0.8}
    \draw[-{Stealth[length=5pt]}, thick, gray!70]
      (gdone.west)
        -- node[above, font=\sffamily\scriptsize] {no} (\outerLoopX, -1.3)
        -- (\outerLoopX, 9.5)
        -- (7, 9.5)
        -- (step.north);
    \node[font=\sffamily\scriptsize, gray!60!black, anchor=south] at (3.5, 9.5)
      {wait for incoming particles};
  \end{scope}

\end{tikzpicture}}
    \caption{Per-time-step main loop of the Monte Carlo manager. After initialization and particle creation (\texttt{preStep}), each rank enters the \texttt{HandleAll} routine (dashed box), which iterates over all locally available particles: each particle is advanced by \texttt{physics.step} and, depending on the outcome, either moves to a local cell, is transferred to a neighbor rank via RDMA (Algorithm~\ref{alg:transfer}), or is removed (with the distributed counter decremented accordingly). The inner loop (right-side arrow) continues until no particles remain in the local handlers. Control then passes to the termination check (Section~\ref{sec:termination}): if all particles across all ranks are globally done, the step concludes with population control and \texttt{postStep}; otherwise, the rank re-enters \texttt{HandleAll} to process any particles that have arrived via RDMA in the interim.}
    \label{fig:main_loop}
\end{figure*}

The architecture is described in terms of one-sided RDMA operations. The per-pair channel concept, aggregation strategy, and termination protocol apply to all backends; the specific SPSC ring-buffer Handler structure described below is used by the one-sided backends (OFI, IBV, and MPI RMA), while the point-to-point backend replaces it with per-peer message queues. Figure~\ref{fig:main_loop} provides an overview of the complete per-time-step loop: each rank initializes its particles, then enters a main loop that repeatedly propagates particles via the physics module. Depending on the outcome, a particle either moves to a local cell, is transferred to a neighbor rank via the selected backend, or is removed (absorbed/escaped). The loop terminates when a distributed particle counter reaches zero and all ranks confirm they hold no remaining work. The subsections that follow detail each component of this architecture.

\subsection{Architecture Overview}\label{sec:architecture_overview}

\subsubsection{Generality of the Communication Layer}
The communication layer is deliberately separated from the representation of the spatial discretization. In the most general case, the manager only requires that the application determines, after advancing a particle, one of the following outcomes: (i) the particle remains local, (ii) the particle must be transferred to a known destination rank, or (iii) the particle is removed. Any additional metadata needed by the destination rank---for example, a destination cell index, a stable cell identifier, a patch identifier, or a particle-container handle---is carried as part of the particle state. Consequently, the Handler design, aggregation buffers, reallocation protocol, and termination detector do not depend on Voronoi cells specifically. The moving-Voronoi remapping procedure described later is one concrete application-specific layer built on top of this general communication substrate.

\subsubsection{Physics-agnostic Architecture}
The communication engine---implemented in \textsc{STORM} as the \texttt{RDMA\-Monte\-Carlo\-Manager} class---is entirely decoupled from the underlying particle physics. All physics-specific behavior is supplied through three abstract interfaces that the user implements.

The first is the \emph{physics} interface, which defines three callback stages invoked by the manager during each time step (see Figure~\ref{fig:main_loop}):
\begin{enumerate}
    \item \textbf{Pre-step} (\texttt{preStep}): create new packets (e.g., thermal or source photons) and performs any per-time-step initialization required by the physics model.
    \item \textbf{Per-particle step} (\texttt{step}): given a single packet, advances it for one (or more) event(s) according to the physics and returns a status code that tells the manager what to do next---continue propagating the packet within the current cell (the particle's state, e.g., direction or energy, may have been updated by the interaction even though it remains in the same cell), move it to a neighboring cell (the manager then decides whether this is a local index update or an RDMA transfer), mark it as finished (e.g., reached the end of the time step), or remove it. The physics may also spawn new packets on the fly (e.g., via inelastic scattering or fluorescence) by appending them to a designated list. Because each packet stores the index of the cell it currently occupies, the physics module can tally energy deposition and other quantities directly during this step.
    \item \textbf{Post-step} (\texttt{postStep}): finalizes per-time-step tallies and updates material state (e.g., deposited energy, temperature) after all packets have been processed.
\end{enumerate}
The second is the \emph{boundary condition} interface. It provides two methods: one is called by the manager whenever a packet reaches the domain boundary, returning a status that determines whether the packet is absorbed, reflected, or escapes; the other may be called by the physics pre-step to inject new source packets from the boundary (e.g., a driven blackbody surface).

The third is the \emph{population-control} interface. After the main transport loop completes and before the physics post-step, the manager passes all surviving packets to the population-control module, which returns a new list in which the total weight in a cell is conserved but the number of packets has been adjusted. Any population-control strategy that satisfies this interface can be used. In our implementation, we use a \emph{comb} algorithm: within each cell, packets whose count exceeds a user-specified threshold $N_{\max}$ are merged by redistributing their combined energy over fewer representative packets, while cells with fewer than a minimum count $N_{\min}$ have their packets split to maintain statistical resolution.

Because the manager interacts with these three components only through their abstract interfaces, \textsc{STORM} can be reused with different physics models (e.g., gray or multigroup photon radiation, neutrino transport), boundary prescriptions, population-control strategies, and spatial discretizations without modification.

\subsection{Handler Design}\label{sec:handler_design}

\subsubsection{Motivation for One-sided Communication}
In a domain-decomposed Monte Carlo particle transport simulation, packets cross rank boundaries at unpredictable times and in unpredictable directions. With classical two-sided communication (\texttt{MPI\_Send}/\texttt{MPI\_Recv}), data movement is expressed through matched sends and receives. The receiver must eventually post a matching receive for each send, using either blocking or nonblocking operations. Nonblocking operations can overlap computation with communication, but the application must often periodically enter MPI to match incoming messages and advance completion. This creates a fundamental tension---a rank that is busy propagating its own particles cannot simultaneously service incoming transfers from all of its neighbors without either (i)~probing or testing for messages from every peer after many particle steps, introducing overhead in the main Monte Carlo loop, or (ii)~batching transfers to synchronization points, which delays packets and underutilizes the interconnect.

One-sided operations eliminate this coupling. With RDMA, the sender writes the particle data directly into the receiver's pre-registered memory; the receiver need not execute any communication call at all. This yields three concrete advantages:
\begin{enumerate}
    \item \textbf{Asynchronous progress.} A rank can continue propagating local particles while incoming packets are deposited into its buffers by remote ranks, without any interruption or polling.
    \item \textbf{Low latency.} Each transfer is a single RDMA operation (e.g., \texttt{fi\_write}) or a \texttt{MPI\_Put} into a known memory offset, avoiding the handshake and envelope matching overhead of two-sided messaging.
    \item \textbf{Decoupled load balance.} Fast ranks that finish their local work early can immediately push particles to slower neighbors; they do not stall waiting for the receiver to post a receive. Conversely, an overloaded rank is never forced to interrupt its computation to participate in a receive call.
\end{enumerate}
That said, modern two-sided implementations can partially mitigate these drawbacks through nonblocking operations, message aggregation, and manual progress loops. Indeed, a well-optimized MPI layer may itself use RDMA internally for large messages---for example, via a rendezvous protocol that pins memory and performs a zero-copy transfer. However, the two-sided semantics are preserved: the receiving rank must still post a matching receive call for the transfer to complete, and the sender must periodically call \texttt{MPI\_Test} or \texttt{MPI\_Wait} to detect completion. The result is not a lack of RDMA in MPI, but a loss of application control over the delicate details of asynchronous progress, remote-buffer layout, completion granularity, and ordering. We therefore also provide an optimized point-to-point backend that employs nonblocking sends, message aggregation, and manual progress loops, offering a competitive alternative when direct low-level RDMA access is unavailable or when memory-pinning constraints make one-sided communication impractical. The per-pair channel concept, aggregation strategy, and termination protocol described below apply uniformly to both the RDMA and point-to-point backends. The specific SPSC ring-buffer Handler structure (Section~\ref{sec:handler_design}) is used by the one-sided backends; the point-to-point backend replaces it with per-peer message queues while preserving the same per-pair isolation guarantees.

\subsubsection{Per-pair Channel Architecture}
The most straightforward RDMA design would give each rank a single particle buffer, exposed as one MPI window, into which any remote rank may write. However, this approach suffers from three serious problems.
First, because multiple senders may target the same receiver concurrently, every write must be protected by a global lock on the receiver's buffer. Under high traffic, this lock forces senders to execute sequentially: they queue behind one another even when they are depositing particles into entirely different slots. Second, the receiver itself needs to read from the same buffer to process incoming particles, so it must contend for the same lock, coupling the receiver's computation rate to the senders' transfer rate. Third---and most critically---the lock ordering between ranks is not naturally acyclic. If rank~$A$ holds the lock on $B$'s buffer while $B$ simultaneously holds the lock on $A$'s buffer, neither can proceed, resulting in deadlock. With $P$ ranks, the number of potential deadlock cycles grows combinatorially, making the design unreliable at scale.

These shortcomings motivate a finer-grained architecture in which each \emph{pair} of ranks has its own private channel, eliminating cross-sender contention and trivially avoiding deadlock by construction.

To exploit these properties, we introduce a dedicated communication channel---termed a \textit{Handler}---for every ordered pair of neighboring ranks in the domain decomposition. If rank~$A$'s subdomain is adjacent to rank~$B$'s, then rank~$B$ allocates a Handler into which $A$ will write, and $A$ allocates a separate Handler for traffic arriving from~$B$. Handlers are instantiated whenever a new neighboring pair appears: at the beginning of the run for the initial decomposition, and after each load-rebalancing repartition for any newly adjacent rank pairs. At creation time, their memory is registered for RDMA access, so remote memory is already accessible when the first particle crosses the subdomain boundary.

\subsubsection{Handler Structure}
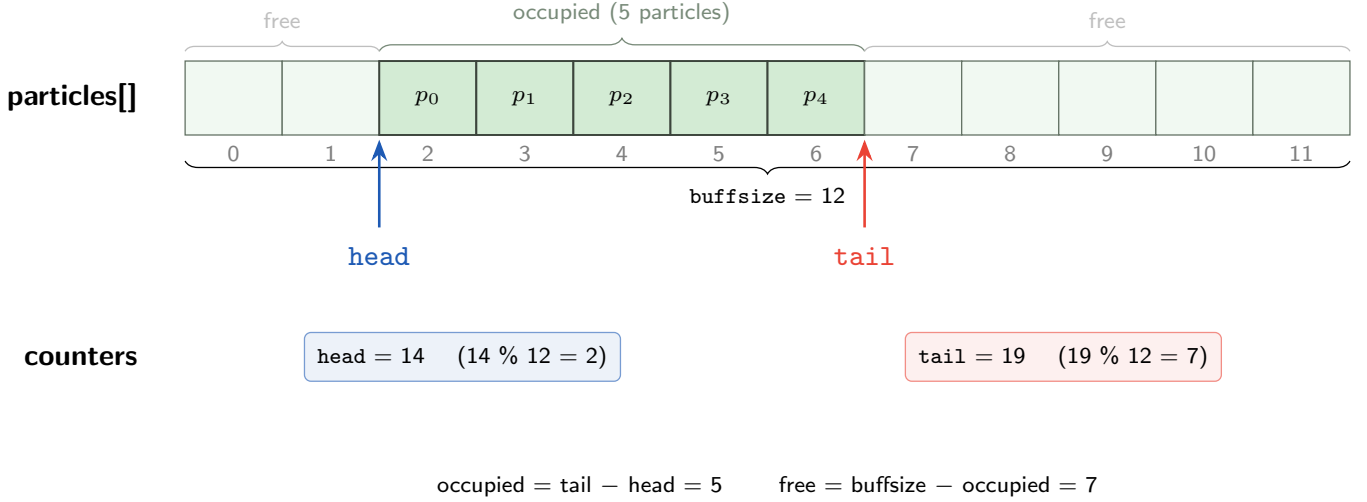
\begin{figure*}
    \centering
    \resizebox{\textwidth}{!}{\begin{tikzpicture}[>=Stealth, font=\sffamily\normalsize]

  \begin{scope}[shift={(1.5, -1.5)}]
    \def\cw{1.05}
    \def\xstart{0}

    \node[font=\sffamily\small\bfseries, anchor=east] at (-0.4, -0.3) {particles[]};

    \foreach \i in {0,...,11} {
      \pgfmathsetmacro{\xl}{\xstart + \i*\cw}
      \pgfmathsetmacro{\xr}{\xl + \cw}
      \pgfmathtruncatemacro{\isOcc}{(\i >= 2) * (\i < 7)}
      \ifnum \isOcc>0
        \fill[partcolor!80, draw=partcolor!30!black, semithick]
          (\xl, -0.7) rectangle (\xr, 0.1);
      \else
        \fill[partcolor!25, draw=partcolor!50!black, thin]
          (\xl, -0.7) rectangle (\xr, 0.1);
      \fi
    }
    \foreach \i/\lab in {0/\phantom{-}, 1/\phantom{-}, 2/$p_0$, 3/$p_1$, 4/$p_2$, 5/$p_3$,
                          6/$p_4$, 7/\phantom{-}, 8/\phantom{-}, 9/\phantom{-}, 10/\phantom{-}, 11/\phantom{-}} {
      \pgfmathsetmacro{\xc}{\xstart + \i*\cw + 0.5*\cw}
      \node[font=\sffamily\scriptsize] at (\xc, -0.3) {\lab};
    }
    \foreach \i in {0,...,11} {
      \pgfmathsetmacro{\xc}{\xstart + \i*\cw + 0.5*\cw}
      \node[font=\sffamily\scriptsize, gray] at (\xc, -0.9) {\i};
    }

    \pgfmathsetmacro{\xleft}{\xstart}
    \pgfmathsetmacro{\xright}{\xstart + 12*\cw}
    \draw[decorate, decoration={brace, amplitude=4pt, mirror, raise=8pt}]
      (\xleft, -0.7) -- (\xright, -0.7)
      node[midway, below=13pt, font=\sffamily\scriptsize] {\texttt{buffsize} = 12};

    \pgfmathsetmacro{\headx}{\xstart + 2*\cw}
    \draw[->, thick, arrowblue] (\headx, -1.7) -- (\headx, -0.75);
    \node[font=\sffamily\small\bfseries, arrowblue, anchor=north] at (\headx, -1.8)
      {\texttt{head}};

    \pgfmathsetmacro{\tailx}{\xstart + 7*\cw}
    \draw[->, thick, rank1] (\tailx, -1.7) -- (\tailx, -0.75);
    \node[font=\sffamily\small\bfseries, rank1, anchor=north] at (\tailx, -1.8)
      {\texttt{tail}};

    \def\cby{-3.3}
    \node[font=\sffamily\small\bfseries, anchor=east] at (-0.4, \cby+0.2) {counters};

    \node[draw=arrowblue!60, fill=arrowblue!8, rounded corners=2pt,
          font=\sffamily\scriptsize, inner sep=4pt, minimum width=2.8cm]
      at (3.0, \cby+0.2) {\texttt{head} = 14 \quad (14 \% 12 = 2)};

    \node[draw=rank1!60, fill=rank1!8, rounded corners=2pt,
          font=\sffamily\scriptsize, inner sep=4pt, minimum width=2.8cm]
      at (9.5, \cby+0.2) {\texttt{tail} = 19 \quad (19 \% 12 = 7)};

    \def\aby{-4.5}
    \node[font=\sffamily\scriptsize, anchor=center] at (6.3, \aby)
      {occupied = tail $-$ head = 5 \qquad free = buffsize $-$ occupied = 7};

    \pgfmathsetmacro{\occLeft}{\xstart + 2*\cw}
    \pgfmathsetmacro{\occRight}{\xstart + 7*\cw}
    \draw[decorate, decoration={brace, amplitude=4pt, raise=3pt}, partcolor!60!black]
      (\occLeft, 0.1) -- (\occRight, 0.1)
      node[midway, above=8pt, font=\sffamily\scriptsize, partcolor!60!black]
      {occupied (5 particles)};

    \pgfmathsetmacro{\freeLeftA}{\xstart}
    \pgfmathsetmacro{\freeRightA}{\xstart + 2*\cw}
    \draw[decorate, decoration={brace, amplitude=3pt, raise=3pt}, gray!50]
      (\freeLeftA, 0.1) -- (\freeRightA, 0.1)
      node[midway, above=7pt, font=\sffamily\scriptsize, gray!50] {free};
    \pgfmathsetmacro{\freeLeftB}{\xstart + 7*\cw}
    \pgfmathsetmacro{\freeRightB}{\xstart + 12*\cw}
    \draw[decorate, decoration={brace, amplitude=3pt, raise=3pt}, gray!50]
      (\freeLeftB, 0.1) -- (\freeRightB, 0.1)
      node[midway, above=7pt, font=\sffamily\scriptsize, gray!50] {free};

  \end{scope}

\end{tikzpicture}}
    \caption{Ring-buffer layout of a single Handler. The \texttt{particles[\,]} array is a circular buffer; dark cells are occupied slots in the region from $\texttt{head} \bmod \texttt{buffsize}$ to $\texttt{tail} \bmod \texttt{buffsize}$. Two RDMA-registered memory regions expose the particle data and the $[\texttt{head}, \texttt{tail}]$ counter pair, respectively. Because each Handler has exactly one sender and one receiver (SPSC), no locks are required on the transfer path.}
    \label{fig:handler_buffer}
\end{figure*}

As illustrated in Figure~\ref{fig:handler_buffer}, each inter-rank Handler is implemented as a single-producer, single-consumer (SPSC) ring buffer backed by two RDMA-accessible memory regions:
\begin{enumerate}
    \item \texttt{particles[\,]}---a pre-allocated circular buffer of particle slots, indexed modulo the buffer capacity \texttt{buffsize}.
    \item \texttt{head} and \texttt{tail}---two monotonically increasing 64-bit counters exposed through a second RDMA-registered region. \texttt{tail} records the total number of particles ever \emph{produced} (written into the buffer by the sender), and \texttt{head} records the total number ever \emph{consumed} (read and retired by the receiver). The occupied count is $\texttt{tail} - \texttt{head}$, and a slot at logical position~$k$ maps to physical index $k \bmod \texttt{buffsize}$.
\end{enumerate}
Because each Handler connects exactly one sender to one receiver, the SPSC invariant is structurally guaranteed: only the sender ever advances \texttt{tail}, and only the receiver ever advances \texttt{head}. This separation of concerns eliminates the need for any remote lock on the transfer path.

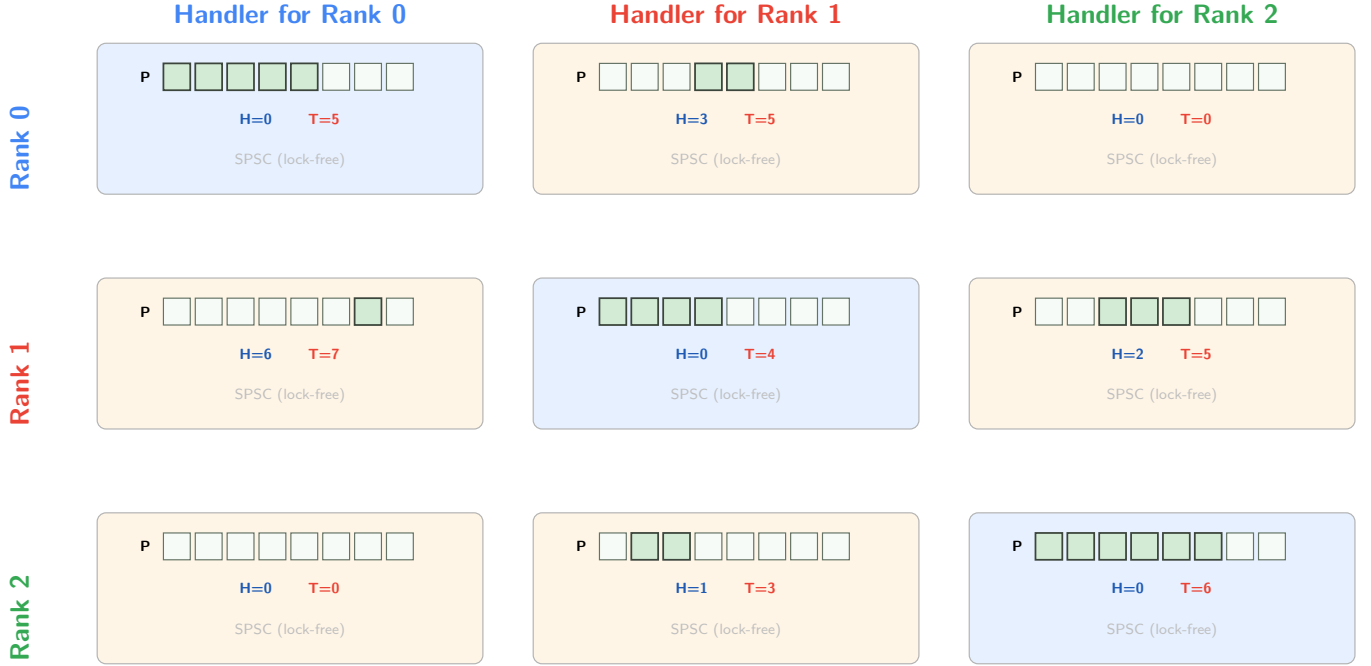
\begin{figure*}
    \centering
    \resizebox{\textwidth}{!}{\begin{tikzpicture}[>=Stealth, font=\sffamily\small]

  \begin{scope}[shift={(0, -5.5)}]
    \node[font=\sffamily\small\bfseries, rank0] at (1.0, 3.6) {Handler for Rank 0};
    \node[font=\sffamily\small\bfseries, rank1] at (6.2, 3.6) {Handler for Rank 1};
    \node[font=\sffamily\small\bfseries, rank2] at (11.4, 3.6) {Handler for Rank 2};

    \node[font=\sffamily\small\bfseries, rank0, rotate=90, anchor=south]
      at (-2.0, 2.0) {Rank 0};
    \node[font=\sffamily\small\bfseries, rank1, rotate=90, anchor=south]
      at (-2.0, -0.8) {Rank 1};
    \node[font=\sffamily\small\bfseries, rank2, rotate=90, anchor=south]
      at (-2.0, -3.6) {Rank 2};

    \drawHandler{1.0}{2.0}  {selfbg}  {}{0}{0}{5}
    \drawHandler{6.2}{2.0}  {remotebg}{}{0}{3}{5}
    \drawHandler{11.4}{2.0} {remotebg}{}{0}{0}{0}
    \drawHandler{1.0}{-0.8} {remotebg}{}{0}{6}{7}
    \drawHandler{6.2}{-0.8} {selfbg}  {}{0}{0}{4}
    \drawHandler{11.4}{-0.8}{remotebg}{}{0}{2}{5}
    \drawHandler{1.0}{-3.6} {remotebg}{}{0}{0}{0}
    \drawHandler{6.2}{-3.6} {remotebg}{}{0}{1}{3}
    \drawHandler{11.4}{-3.6}{selfbg}  {}{0}{0}{6}
  \end{scope}
\end{tikzpicture}}
    \caption{The \texttt{rankHandlers[\,]} matrix for a three-rank communicator. Each cell $(i,j)$ is the Handler that rank~$i$ holds for rank~$j$. Diagonal cells (blue background) are self-handlers using plain allocated memory. Off-diagonal cells (orange background) are remote handlers backed by RDMA-registered memory. Each Handler is an SPSC ring buffer: \textbf{P} shows the particle buffer with dark cells indicating occupied slots, and the labels \textbf{H} and \textbf{T} denote the current \texttt{head} and \texttt{tail} counters. All handlers are lock-free.}
    \label{fig:handler_matrix}
\end{figure*}

\begin{figure*}
    \centering
    \resizebox{0.82\textwidth}{!}{\begin{tikzpicture}[>=Stealth, font=\sffamily\small,
    cell/.style={minimum width=0.72cm, minimum height=0.52cm,
                 draw, thin, font=\sffamily\scriptsize, inner sep=0pt},
    occcell/.style={cell, fill=partcolor!50, draw=partcolor!70!black},
    freecell/.style={cell, fill=white, draw=gray!30},
    targetcell/.style={cell, fill=green!50!black!40, draw=green!70!black, thick},
    lbox/.style={draw, thick, fill=#1, rounded corners=2pt,
                 font=\sffamily\scriptsize, inner sep=3pt,
                 minimum width=1.0cm, minimum height=0.4cm},
    snum/.style={shape=circle, draw, thick, inner sep=0pt,
                 minimum size=11pt, font=\sffamily\tiny\bfseries},
  ]

  \node[font=\sffamily\bfseries, rank0] at (2.0, 7.55) {Rank A (sender)};
  \draw[rank0!40, thick, rounded corners=6pt, fill=selfbg]
    (0, -0.1) rectangle (4.0, 7.2);

  \node[draw=rank0!60, fill=rank0!8, thick, rounded corners=4pt,
        minimum width=2.6cm, minimum height=0.7cm,
        font=\sffamily\small] (pA) at (2.0, 6.65) {Particles $p_0\!,\ldots,p_{N-1}$};

  \def\sx{0.5}
  \def\slx{0.9}
  \def\ya{5.4}   
  \def\yb{3.6}   
  \def\yc{1.6}   

  \node[snum] at (\sx, \ya) {1};
  \node[font=\sffamily\scriptsize, anchor=west] at (\slx, \ya)
    {Read remote counters};
  \node[font=\sffamily\tiny, anchor=west, gray!50!black] at (0.9, 4.95)
    {if full $\to$ reallocation};

  \node[snum] at (\sx, \yb) {2};
  \node[font=\sffamily\scriptsize, anchor=west] at (\slx, \yb)
    {Write particles at tail};
  \node[font=\sffamily\tiny, anchor=west, gray!50!black] at (0.9, 3.15)
    {wraps if needed};

  \node[snum] at (\sx, \yc) {3};
  \node[font=\sffamily\scriptsize, anchor=west] at (\slx, \yc)
    {Atomic \texttt{FetchAndAdd}};
  \node[font=\sffamily\tiny, anchor=west, gray!50!black] at (0.9, 1.15)
    {publishes particles};

  \node[font=\sffamily\tiny, gray!50, inner sep=2pt] at (2.0, 0.35)
    {SPSC: no mutex required};

  \node[font=\sffamily\bfseries, rank1] at (10.8, 7.55)
    {Rank B's Handler (\emph{passive})};
  \draw[rank1!40, thick, rounded corners=6pt, fill=remotebg]
    (5.8, -0.1) rectangle (15.8, 7.2);

  \node[lbox=arrowblue!15, anchor=west]
    (headbox) at (6.2, \ya+0.22) {\texttt{head}: 12};
  \node[lbox=rank1!15, anchor=west]
    (tailbox) at (6.2, \ya-0.22) {\texttt{tail}: 15};

  \def\px{7.3}
  \node[font=\sffamily\scriptsize\bfseries, anchor=south west] at (\px-0.36, \yb+0.55)
    {\texttt{particles[\,]} (ring buffer)};
  \node[occcell]    (s0) at (\px+0*0.76, \yb) {};
  \node[freecell]   (s1) at (\px+1*0.76, \yb) {};
  \node[freecell]   (s2) at (\px+2*0.76, \yb) {};
  \node[freecell]   (s3) at (\px+3*0.76, \yb) {};
  \node[occcell]    (s4) at (\px+4*0.76, \yb) {$p_a$};
  \node[occcell]    (s5) at (\px+5*0.76, \yb) {$p_b$};
  \node[occcell]    (s6) at (\px+6*0.76, \yb) {$p_c$};
  \node[targetcell] (s7) at (\px+7*0.76, \yb) {\color{black!80!green}$p_0$};
  \node[targetcell] (s0new) at (\px+0*0.76, \yb) {\color{black!80!green}$p_1$};
  \foreach \i in {0,...,7} {
    \node[font=\sffamily\tiny, gray!60] at (\px+\i*0.76, \yb-0.42) {\i};
  }
  \pgfmathsetmacro{\headslot}{\px + 4*0.76}
  \pgfmathsetmacro{\tailslot}{\px + 7*0.76}
  \draw[->, thick, arrowblue] (\headslot, \yb-0.7) -- (\headslot, \yb-0.42);
  \node[font=\sffamily\tiny\bfseries, arrowblue] at (\headslot, \yb-0.85) {H};
  \draw[->, thick, rank1] (\tailslot, \yb-0.7) -- (\tailslot, \yb-0.42);
  \node[font=\sffamily\tiny\bfseries, rank1] at (\tailslot, \yb-0.85) {T};
  \node[font=\sffamily\tiny, gray!50, anchor=west] at (\px+8.2*0.76, \yb)
    {wraps to slot 0};

  \node[lbox=rank1!15, anchor=west]
    (tailUpdate) at (6.2, \yc) {\texttt{tail}: $15 \!\to\! 17$};
  \node[font=\sffamily\tiny, gray!50, anchor=west] at (9.5, \yc)
    {atomic FetchAndAdd($+2$)};

  \def\al{4.0}

  \draw[{Stealth[length=4pt]}-, arrowblue, thick, densely dashed]
    (\al, \ya+0.12) -- (headbox.west);
  \draw[{Stealth[length=4pt]}-, arrowblue, thick, densely dashed]
    (\al, \ya-0.12) -- (tailbox.west);

  \draw[-{Stealth[length=4pt]}, rank2!80!black, thick]
    (\al, \yb+0.15) .. controls (5.5, \yb+0.8) and (11.5, \yb+0.65) .. (s7.north);
  \draw[-{Stealth[length=4pt]}, rank2!80!black, thick]
    (\al, \yb-0.15) .. controls (5.5, \yb-0.8) and (6.5, \yb-0.65) .. (s0new.south);

  \draw[-{Stealth[length=4pt]}, rank2!70!black, thick, dotted]
    (\al, \yc) to[bend right=8] (tailUpdate.west);

  \begin{scope}[shift={(12.2, 0.5)}]
    \node[font=\sffamily\tiny\bfseries, anchor=west] at (0, 0.65) {Legend:};
    \draw[{Stealth[length=3pt]}-, arrowblue, thick, densely dashed]
      (0, 0.25) -- (0.7, 0.25);
    \node[font=\sffamily\tiny, anchor=west] at (0.8, 0.25) {read (one-sided)};
    \draw[-{Stealth[length=3pt]}, rank2!80!black, thick]
      (0, -0.1) -- (0.7, -0.1);
    \node[font=\sffamily\tiny, anchor=west] at (0.8, -0.1) {write (one-sided)};
    \draw[-{Stealth[length=3pt]}, rank2!70!black, thick, dotted]
      (0, -0.45) -- (0.7, -0.45);
    \node[font=\sffamily\tiny, anchor=west] at (0.8, -0.45) {atomic (FetchAndAdd)};
  \end{scope}

\end{tikzpicture}}
    \caption{Lock-free one-sided RDMA transfer of particles from rank~$A$ to rank~$B$'s Handler. Rank~$B$ executes no code during the entire operation. Dashed blue arrows denote one-sided reads (data flows from $B$'s memory to $A$); solid green arrows denote one-sided writes ($A$ deposits data into $B$'s memory). The sender reads the remote $[\texttt{head}, \texttt{tail}]$ counters~(1), writes the particles contiguously into the ring buffer at the tail position~(2), and atomically increments the remote \texttt{tail} via \texttt{FetchAndAdd}~(3). No lock is required because each Handler has exactly one sender (SPSC). Highlighted cells show the newly written data. If the buffer is full at step~(1), the reallocation protocol (Section~\ref{sec:reallocation}) is invoked before retrying.}
    \label{fig:rdma_procedures}
\end{figure*}
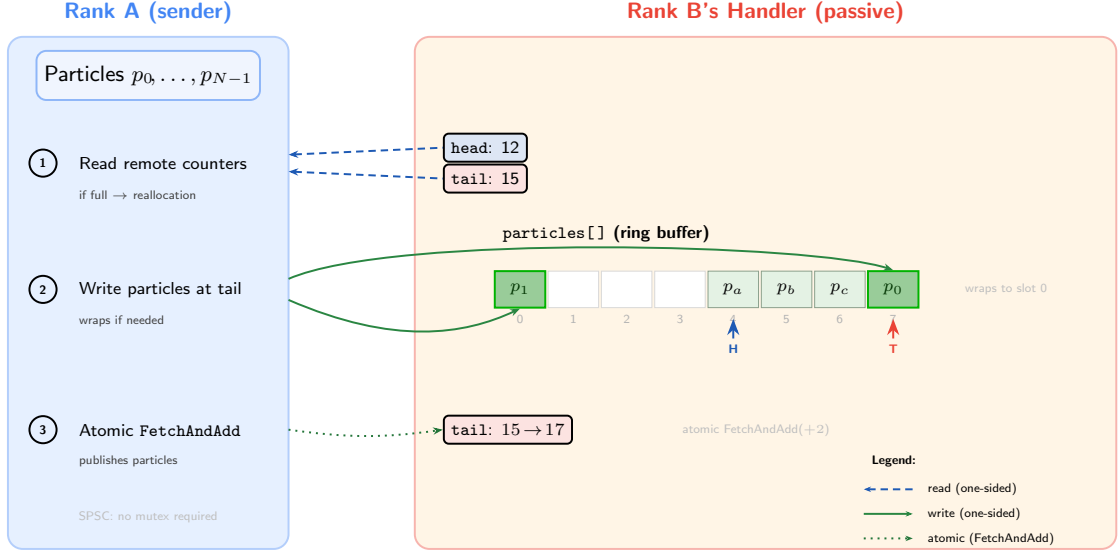

Figure~\ref{fig:handler_buffer} shows the detailed ring-buffer layout of a single Handler: particles occupy contiguous slots from $\texttt{head} \bmod \texttt{buffsize}$ to $\texttt{tail} \bmod \texttt{buffsize}$, wrapping around at the end of the array when necessary.
Figure~\ref{fig:handler_matrix} shows the full Handler matrix for a three-rank example: each row corresponds to a rank's local view, each column to a peer. Off-diagonal entries are remote handlers backed by RDMA-registered memory and carry all inter-rank traffic. Diagonal entries are \emph{self-handlers}: each rank also holds a Handler addressed to itself, backed by plain allocated memory with no RDMA exposure. These store the census particles from the previous time step, which form the initial population at the start of each step; no communication is performed through them. They share the same ring-buffer layout as remote handlers, so the transport loop can dequeue local and remote particles uniformly, but their purpose is local bookkeeping rather than transfer. All handlers are lock-free on the transfer path.

The ring-buffer layout provides two key advantages over an alternative design based on separate index arrays. First, particles are always stored contiguously (modulo wrap-around), so the sender writes them in bulk starting at the tail position with no scattered index lookups---a significant throughput improvement over writing individual particles into random free slots. Second, and more critically, the SPSC structure eliminates all remote locking from the transfer path: the sender only ever increments \texttt{tail} (via an atomic \texttt{FetchAndAdd}), and the receiver only ever increments \texttt{head}. Neither side touches the other's counter during normal operation. The receiver retires processed particles by copying the occupied region to a local list and advancing \texttt{head}---an operation that touches only local memory and requires no remote coordination.

\subsection{Transferring Particles}\label{sec:transfer}
The transfer protocol is fully one-sided and lock-free: the receiver (rank~$B$) remains completely passive while the sender (rank~$A$) carries out all memory operations via RDMA without acquiring any locks. Algorithm~\ref{alg:transfer} gives the pseudocode, and Figure~\ref{fig:rdma_procedures} illustrates the corresponding data flow.

The sender first reads the remote $[\texttt{head}, \texttt{tail}]$ counters to determine available capacity. Let $N_p$ denote the number of particles to be transferred. If the ring buffer is full---i.e., $\texttt{tail} - \texttt{head} + N_p > \texttt{buffsize}$---the reallocation protocol (Section~\ref{sec:reallocation}) is invoked, and the counters are re-read. Once sufficient capacity is confirmed, the sender writes all $N_p$ particles contiguously into the ring buffer starting at position $\texttt{tail} \bmod \texttt{buffsize}$, issuing a second \texttt{Put} for the wrap-around segment if the write crosses the end of the array. Finally, the sender atomically increments the remote \texttt{tail} counter by $N_p$ via a single \texttt{FetchAndAdd} operation, which simultaneously publishes the new particles to the receiver and returns the old counter value as a consistency check.

Because every memory access targets rank~$B$'s window, rank~$B$ issues no explicit receive; it discovers newly arrived particles simply by observing that its local $\texttt{tail} - \texttt{head}$ has increased. No lock is required at any point: the SPSC invariant guarantees that only rank~$A$ writes \texttt{tail} and only rank~$B$ writes \texttt{head}, so the two sides never contend on the same counter. The atomic \texttt{FetchAndAdd} ensures that the tail increment is visible to the receiver only after all particle data has been written.

\begin{algorithm}[t]
\DontPrintSemicolon
\SetKwInOut{Input}{Input}
\SetKwFunction{Get}{Get}
\SetKwFunction{Put}{Put}
\SetKwFunction{FetchAndAdd}{FetchAndAdd}
\SetKwFunction{RequestRealloc}{RequestReallocation}
\Input{$N_p$ particles to transfer; remote Handler $H_B$ on rank $B$}
\BlankLine
$[h, t] \leftarrow$ \Get{$H_B$.\textnormal{counters}} \tcp*{read remote [head, tail]}
\While{$t - h + N_p > H_B.\textnormal{buffsize}$}{
    \RequestRealloc{$B$}\;
    $[h, t] \leftarrow$ \Get{$H_B$.\textnormal{counters}}\;
}
$s \leftarrow t \bmod \textnormal{buffsize}$\;
$\textnormal{first} \leftarrow \min(N_p, \textnormal{buffsize} - s)$\;
\Put{particles$[0\,..\,\textnormal{first}) \to H_B$.\textnormal{particles}$[s\,..\,s{+}\textnormal{first})$}\;
\If{$\textnormal{first} < N_p$}{
    \Put{particles$[\textnormal{first}\,..\,N_p) \to H_B$.\textnormal{particles}$[0\,..\,N_p{-}\textnormal{first})$} \tcp*{wrap-around}
}
\FetchAndAdd{$N_p \to H_B$.\textnormal{tail}} \tcp*{atomically publish}
\caption{Lock-free one-sided RDMA particle transfer from rank $A$ to rank $B$}\label{alg:transfer}
\end{algorithm}

\subsection{Particle Aggregation}\label{sec:aggregation}
Although the lock-free transfer protocol (Algorithm~\ref{alg:transfer}) already handles batches of $N_p$ particles in a single invocation, each call still requires reading the remote counters, issuing one or two RDMA \texttt{Put} operations, and performing an atomic \texttt{FetchAndAdd}. When a rank has many outgoing particles destined for the same peer---a common occurrence, since neighboring subdomains often exchange bursts of particles across a shared boundary---invoking the protocol once per particle incurs overhead from repeated RDMA round-trips.

To mitigate this, each rank maintains a local \emph{aggregation buffer} per peer. Outgoing particles are not transferred immediately; instead, they are appended to the aggregation buffer for the target rank. When the buffer reaches a configurable threshold size, or when the rank's local work queue is exhausted, the entire buffer is flushed in a single batched transfer. The batched transfer reads the remote counters once, writes all particle data contiguously into the ring buffer, and publishes the batch with a single \texttt{FetchAndAdd}---amortizing the cost of one counter read and one atomic update over the entire batch.

This aggregation reduces the total number of RDMA initiation round-trips and atomic operations by a factor proportional to the average batch size. In communication-intensive regimes where hundreds of particles per time step cross each domain boundary, the reduction in overhead is substantial. The aggregation threshold is a tunable parameter; in our benchmarks, we use a default batch size of 64~particles, which provides a good balance between latency (particles are not held in the buffer indefinitely) and throughput (the per-particle RDMA overhead is effectively negligible). Aggregation is equally applicable to the point-to-point backend, where it reduces the number of individual \texttt{MPI\_Isend} calls.

\subsection{Buffer Reallocation}\label{sec:reallocation}
Each Handler's particle buffer is pre-allocated with a fixed capacity. During the simulation, the traffic between two ranks may exceed this initial size, so the buffer must be enlarged at runtime. Reallocation is triggered when a sender discovers, during the transfer protocol (Algorithm~\ref{alg:transfer}), that the receiver's ring buffer is full---i.e., $\texttt{tail} - \texttt{head} + N_p > \texttt{buffsize}$. Enlarging a buffer requires deregistering the old RDMA-exposed memory and registering a new, larger region, and the sender must obtain the receiver's updated remote memory key and base address before it can resume transfers.

\paragraph{Asynchronous reallocation (one-sided backends).}
The OFI and IBV backends use a fully asynchronous protocol that allows computation to overlap with reallocation. When a sender discovers that its peer's buffer is full, it fires a nonblocking reallocation request (containing the requested growth factor) to the peer via \texttt{MPI\_Isend} and \emph{returns immediately without blocking}. The particles that could not be transferred remain in the local send buffer and are retried on subsequent flush attempts.

On the receiver side, a persistent listener polls for incoming asynchronous reallocation requests. When a request arrives, the receiver enlarges the local Handler's buffer unilaterally---allocating a new, larger RDMA-registered memory region, copying existing contents, and deregistering the old region---then sends back a metadata message containing the updated remote memory key, base address, and new buffer capacity. This metadata is sent via a nonblocking \texttt{MPI\_Isend}, so the receiver also never blocks.

The sender periodically calls a progress function that performs three operations (each at a configurable polling frequency):
\begin{enumerate}
    \item \textbf{Progress outgoing sends:} polls pending factor-send and metadata-send requests for completion.
    \item \textbf{Check metadata updates:} probes for incoming metadata responses from peers. When a response arrives, the sender updates its local copy of the peer's remote addresses and keys, enabling future transfers to use the newly enlarged buffer.
    \item \textbf{Handle incoming requests:} serves incoming reallocation requests from other ranks that need \emph{this} rank to enlarge a buffer, performing the local reallocation and replying with metadata.
\end{enumerate}
While a reallocation for a given peer is pending, the sender skips flushes to that peer and queues the particles for later delivery. This ensures that no rank ever blocks waiting for a reallocation to complete---the main transport loop continues processing other particles and servicing other peers. Deadlock is impossible because no rank ever enters a blocking wait: the protocol is entirely nonblocking and progress is guaranteed by the periodic polling.

Algorithms~\ref{alg:async_realloc_sender} and~\ref{alg:async_realloc_progress} summarize the sender- and receiver-side logic, respectively.

\begin{algorithm}[t]
\DontPrintSemicolon
\SetKwInOut{Input}{Input}
\SetKwFunction{Isend}{MPI\_Isend}
\Input{Peer rank $b$ whose buffer is full; growth factor $f$}
\BlankLine
\If{$b \in \texttt{pendingRanks}$}{\Return \tcp*{already in flight}}
Add $b$ to \texttt{pendingRanks}\;
\Isend{$f$, $b$, \textsc{AsyncRequest}} \tcp*{nonblocking}
\Return \tcp*{resume processing}
\caption{Asynchronous reallocation: sender initiates a request. The sender does not block; particles for peer~$b$ are queued until the reallocation completes.}\label{alg:async_realloc_sender}
\end{algorithm}

\begin{algorithm}[t]
\DontPrintSemicolon
\SetKwFunction{ProgressSends}{ProgressOutgoingSends}
\SetKwFunction{CheckMeta}{CheckMetadataUpdates}
\SetKwFunction{HandleInc}{HandleIncomingRequests}
\SetKwFunction{LocalRealloc}{LocalReallocate}
\SetKwFunction{UpdateRemote}{UpdatePeerRemoteInfo}
\SetKwFunction{Isend}{MPI\_Isend}
\BlankLine
\tcp{(a) Poll pending sends for completion.}
\ProgressSends{}\;
\BlankLine
\tcp{(b) Check for metadata from peers.}
\ForEach{completed metadata message from rank~$r$}{
    \UpdateRemote{$r$, metadata} \tcp*{new rkey, addr, size}
    Remove $r$ from \texttt{pendingRanks}\;
}
\BlankLine
\tcp{(c) Serve incoming realloc requests.}
\ForEach{incoming request from rank~$r$ with factor~$f$}{
    metadata $\leftarrow$ \LocalRealloc{$r$, $f$} \tcp*{enlarge local buffer}
    \Isend{metadata, $r$, \textsc{Metadata}} \tcp*{nonblocking reply}
}
\caption{Asynchronous reallocation: periodic progress. Called by every rank during the main transport loop. Each sub-step has a configurable polling frequency to amortize overhead}\label{alg:async_realloc_progress}
\end{algorithm}

\paragraph{Synchronous reallocation (fallback backends).}
The MPI RMA and point-to-point backends use a synchronous protocol in which the sender blocks until the reallocation completes. To avoid deadlock, each rank maintains a persistent listener for incoming reallocation requests and continues to poll and serve them while waiting for its own acknowledgment. A wait-for priority rule breaks potential cycles: if a rank is waiting for rank~$x$ and $x$'s own request arrives, that request is served first. As a further optimization, before sending a new request, the rank drains all pending incoming requests; if the target peer's request is among them, it is served immediately, and the outgoing request is skipped.

\paragraph{Buffer recreation.}
In both protocols, buffer recreation follows the same steps: a new RDMA-registered region is allocated with the requested capacity, the contents of the old buffer are copied, and the old region is deregistered. In the asynchronous protocol, the receiver performs this locally and sends back the new metadata; in the synchronous protocol, both ranks enter a coordinated rendezvous to exchange updated buffer sizes and remote memory keys.

\subsection{Termination}\label{sec:termination}
A Monte Carlo step is complete only when every particle packet across all ranks has finished its time step. Detecting this global condition in a distributed-memory setting is nontrivial: a rank that has locally exhausted its particles cannot conclude that the simulation is finished, because packets may still be in flight toward it from other ranks.

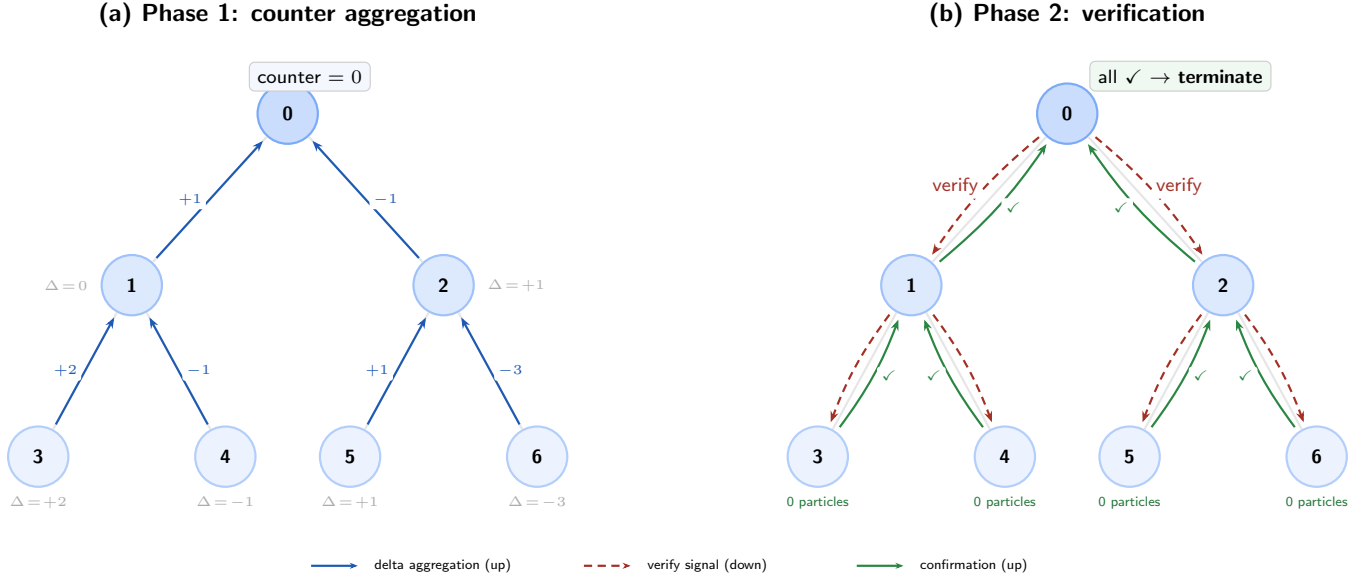
\begin{figure*}
    \centering
    \resizebox{\textwidth}{!}{\begin{tikzpicture}[>=Stealth, font=\sffamily\small,
    rnode/.style={circle, draw, thick, minimum size=22pt,
                  font=\sffamily\scriptsize\bfseries},
    leaf/.style={rnode, fill=rank0!10, draw=rank0!40},
    interior/.style={rnode, fill=rank0!18, draw=rank0!50},
    rootn/.style={rnode, fill=rank0!28, draw=rank0!70},
    dlbl/.style={font=\sffamily\tiny, fill=white, inner sep=1.5pt,
                 rounded corners=1pt},
    annot/.style={font=\sffamily\scriptsize, fill=white, draw=gray!40,
                  rounded corners=2pt, inner sep=3pt},
  ]

  \node[font=\sffamily\bfseries, anchor=south] at (3.5, 6.0)
    {(a) Phase 1: counter aggregation};

  \node[rootn]    (a0) at (3.5, 5.0) {0};
  \node[interior] (a1) at (1.5, 2.8) {1};
  \node[interior] (a2) at (5.5, 2.8) {2};
  \node[leaf]     (a3) at (0.3, 0.6) {3};
  \node[leaf]     (a4) at (2.7, 0.6) {4};
  \node[leaf]     (a5) at (4.3, 0.6) {5};
  \node[leaf]     (a6) at (6.7, 0.6) {6};

  \foreach \p/\c in {a0/a1, a0/a2, a1/a3, a1/a4, a2/a5, a2/a6} {
    \draw[gray!20, thick] (\p) -- (\c);
  }

  \node[font=\sffamily\tiny, gray!65, anchor=north] at (a3.south)
    {$\Delta\!=\!+2$};
  \node[font=\sffamily\tiny, gray!65, anchor=north] at (a4.south)
    {$\Delta\!=\!-1$};
  \node[font=\sffamily\tiny, gray!65, anchor=north] at (a5.south)
    {$\Delta\!=\!+1$};
  \node[font=\sffamily\tiny, gray!65, anchor=north] at (a6.south)
    {$\Delta\!=\!-3$};

  \node[font=\sffamily\tiny, gray!65, anchor=east] at (a1.west)
    {$\Delta\!=\!0$\;};
  \node[font=\sffamily\tiny, gray!65, anchor=west] at (a2.east)
    {\;$\Delta\!=\!+1$};

  \draw[-{Stealth[length=4pt]}, arrowblue, thick,
        shorten >=2pt, shorten <=2pt]
    (a3) -- (a1) node[dlbl, pos=0.5, left=1pt] {\color{arrowblue}$+2$};
  \draw[-{Stealth[length=4pt]}, arrowblue, thick,
        shorten >=2pt, shorten <=2pt]
    (a4) -- (a1) node[dlbl, pos=0.5, right=1pt] {\color{arrowblue}$-1$};
  \draw[-{Stealth[length=4pt]}, arrowblue, thick,
        shorten >=2pt, shorten <=2pt]
    (a5) -- (a2) node[dlbl, pos=0.5, left=1pt] {\color{arrowblue}$+1$};
  \draw[-{Stealth[length=4pt]}, arrowblue, thick,
        shorten >=2pt, shorten <=2pt]
    (a6) -- (a2) node[dlbl, pos=0.5, right=1pt] {\color{arrowblue}$-3$};
  \draw[-{Stealth[length=4pt]}, arrowblue, thick,
        shorten >=2pt, shorten <=2pt]
    (a1) -- (a0) node[dlbl, pos=0.5, left=1pt] {\color{arrowblue}$+1$};
  \draw[-{Stealth[length=4pt]}, arrowblue, thick,
        shorten >=2pt, shorten <=2pt]
    (a2) -- (a0) node[dlbl, pos=0.5, right=1pt] {\color{arrowblue}$-1$};

  \node[annot, anchor=south, fill=rank0!6] at (a0.north east)
    {counter $= 0$};

  \def\dx{10}
  \node[font=\sffamily\bfseries, anchor=south] at (\dx+3.5, 6.0)
    {(b) Phase 2: verification};

  \node[rootn]    (b0) at (\dx+3.5, 5.0) {0};
  \node[interior] (b1) at (\dx+1.5, 2.8) {1};
  \node[interior] (b2) at (\dx+5.5, 2.8) {2};
  \node[leaf]     (b3) at (\dx+0.3, 0.6) {3};
  \node[leaf]     (b4) at (\dx+2.7, 0.6) {4};
  \node[leaf]     (b5) at (\dx+4.3, 0.6) {5};
  \node[leaf]     (b6) at (\dx+6.7, 0.6) {6};

  \foreach \p/\c in {b0/b1, b0/b2, b1/b3, b1/b4, b2/b5, b2/b6} {
    \draw[gray!20, thick] (\p) -- (\c);
  }

  \draw[-{Stealth[length=4pt]}, rank1!70!black, thick, densely dashed,
        shorten >=2pt, shorten <=2pt]
    (b0) to[bend right=8] node[dlbl, pos=0.45, left=1pt,
      text=rank1!70!black] {\scriptsize verify} (b1);
  \draw[-{Stealth[length=4pt]}, rank1!70!black, thick, densely dashed,
        shorten >=2pt, shorten <=2pt]
    (b0) to[bend left=8] node[dlbl, pos=0.45, right=1pt,
      text=rank1!70!black] {\scriptsize verify} (b2);
  \draw[-{Stealth[length=4pt]}, rank1!70!black, thick, densely dashed,
        shorten >=2pt, shorten <=2pt]
    (b1) to[bend right=8] (b3);
  \draw[-{Stealth[length=4pt]}, rank1!70!black, thick, densely dashed,
        shorten >=2pt, shorten <=2pt]
    (b1) to[bend left=8] (b4);
  \draw[-{Stealth[length=4pt]}, rank1!70!black, thick, densely dashed,
        shorten >=2pt, shorten <=2pt]
    (b2) to[bend right=8] (b5);
  \draw[-{Stealth[length=4pt]}, rank1!70!black, thick, densely dashed,
        shorten >=2pt, shorten <=2pt]
    (b2) to[bend left=8] (b6);

  \draw[-{Stealth[length=4pt]}, rank2!80!black, thick,
        shorten >=2pt, shorten <=2pt]
    (b3) to[bend right=8] node[dlbl, pos=0.5, right=1pt,
      text=rank2!80!black] {$\checkmark$} (b1);
  \draw[-{Stealth[length=4pt]}, rank2!80!black, thick,
        shorten >=2pt, shorten <=2pt]
    (b4) to[bend left=8] node[dlbl, pos=0.5, left=1pt,
      text=rank2!80!black] {$\checkmark$} (b1);
  \draw[-{Stealth[length=4pt]}, rank2!80!black, thick,
        shorten >=2pt, shorten <=2pt]
    (b5) to[bend right=8] node[dlbl, pos=0.5, right=1pt,
      text=rank2!80!black] {$\checkmark$} (b2);
  \draw[-{Stealth[length=4pt]}, rank2!80!black, thick,
        shorten >=2pt, shorten <=2pt]
    (b6) to[bend left=8] node[dlbl, pos=0.5, left=1pt,
      text=rank2!80!black] {$\checkmark$} (b2);
  \draw[-{Stealth[length=4pt]}, rank2!80!black, thick,
        shorten >=2pt, shorten <=2pt]
    (b1) to[bend right=8] node[dlbl, pos=0.5, right=1pt,
      text=rank2!80!black] {$\checkmark$} (b0);
  \draw[-{Stealth[length=4pt]}, rank2!80!black, thick,
        shorten >=2pt, shorten <=2pt]
    (b2) to[bend left=8] node[dlbl, pos=0.5, left=1pt,
      text=rank2!80!black] {$\checkmark$} (b0);

  \node[annot, anchor=south west, fill=rank2!8] at (b0.north east)
    {all $\checkmark$ $\to$ \textbf{terminate}};

  \node[font=\sffamily\tiny, rank2!70!black, anchor=north] at (b3.south)
    {0 particles};
  \node[font=\sffamily\tiny, rank2!70!black, anchor=north] at (b4.south)
    {0 particles};
  \node[font=\sffamily\tiny, rank2!70!black, anchor=north] at (b5.south)
    {0 particles};
  \node[font=\sffamily\tiny, rank2!70!black, anchor=north] at (b6.south)
    {0 particles};

  \begin{scope}[shift={(3.8, -0.8)}]
    \draw[-{Stealth[length=3pt]}, arrowblue, thick]
      (0,0) -- (0.6,0);
    \node[font=\sffamily\tiny, anchor=west] at (0.7,0)
      {delta aggregation (up)};
    \draw[-{Stealth[length=3pt]}, rank1!70!black, thick, densely dashed]
      (3.5,0) -- (4.1,0);
    \node[font=\sffamily\tiny, anchor=west] at (4.2,0)
      {verify signal (down)};
    \draw[-{Stealth[length=3pt]}, rank2!80!black, thick]
      (7,0) -- (7.6,0);
    \node[font=\sffamily\tiny, anchor=west] at (7.7,0)
      {confirmation (up)};
  \end{scope}

\end{tikzpicture}}
    \caption{Tree-based distributed termination detection for $P=7$ ranks arranged in a complete binary tree. (a)~Phase~1: each rank accumulates a local particle-count delta ($\Delta$) and periodically sends it to its parent; interior nodes aggregate their children's deltas with their own before forwarding. When the root's global counter reaches zero, Phase~2 is triggered. (b)~Phase~2: the root broadcasts a \emph{verify} signal down the tree (dashed red arrows). Every rank inspects its local state and reports whether it holds zero particles and has no outstanding transfers. Confirmations propagate back up (solid green arrows). If all ranks confirm, the simulation terminates; otherwise, the counter continues to be tracked.}
    \label{fig:termination}
\end{figure*}

We adopt the tree-based completion-detection algorithm of \citet{Brunner2009}, which proved to be efficient and scalable.
The $P$ ranks are arranged in a complete binary tree rooted at rank~$0$, with rank~$i$ having parent $\lfloor(i-1)/2\rfloor$ and children $2i+1$, $2i+2$ (where they exist). The algorithm proceeds in two phases, also demonstrated in Figure~\ref{fig:termination}.

\paragraph{Phase 1: Counter aggregation.}
A global particle counter is maintained by reducing local changes up the tree. Each rank accumulates a local delta that records net particle creation (e.g., newly created packets), destruction (absorption or escape), and reaching census (time-step termination). Periodically, each nonroot rank sends its accumulated delta to its parent via an \texttt{MPI\_Send} and resets its local accumulator. Interior nodes add incoming deltas from their children to their own accumulator before forwarding up. The root adds all incoming contributions to a global counter.

\paragraph{Phase 2: Verification.}
A zero global counter does not, by itself, guarantee that no particles remain in the system. Because the counter aggregates \emph{net} changes asynchronously, a transient zero can occur when the creation of a new particle on one rank coincides with the removal of an unrelated particle on another: the two deltas cancel before reaching the root, even though a live particle still exists. To guard against such false zeros, the root initiates a verification sweep whenever the counter reaches zero. The root sends a \emph{verify} signal down the tree; each interior node forwards it to its children. On receiving this signal, every rank inspects its own state, confirming that it holds no local particles and has no outstanding outgoing transfers. The results are combined with a global logical AND. If all ranks report true, the simulation is declared complete; otherwise, the counter continues to be tracked, and the verification is retried when the root next sees zero.

The tree topology reduces the communication load on the root from $O\left(P\right)$ to $O\left(\log P\right)$ messages per aggregation round, which is essential for scaling to large rank counts where frequent global reductions would serialize progress.

\subsection{Moving Mesh and Dynamic Decomposition}\label{sec:moving_mesh}
This subsection is specific to the moving-Voronoi application that motivates our implementation. It describes how the general communication scheme is coupled to a hydrodynamics code whose mesh points move and whose domain decomposition may change over time. The communication scheme itself does not require a Voronoi mesh or a moving mesh; only the remapping logic in this subsection is application-specific.

Our target application couples particle transport to a moving, unstructured Voronoi mesh, whose parallel construction is described in \citet{Mizrachi2025madvoro}. The hydrodynamic step advects the mesh-generating points and may also change the domain decomposition to maintain load balance. Both operations rebuild the Voronoi tessellation and can change (i)~the local indexing of cells on each rank, (ii)~the owner rank of a given physical region, and (iii)~the mesh topology itself, since the Voronoi cells deform as their generating points move and new cells may be introduced at runtime through adaptive mesh refinement (AMR). Consequently, particle packets may require migration between ranks even if the Monte Carlo transport operator is not applied.

The central difficulty is that on an unstructured Voronoi mesh, one cannot determine the \emph{exact} owner rank of an arbitrary point by a simple arithmetic mapping (as is possible for regular grids). At best, a rank can compute an approximate owner using a coarse geometric partition (e.g., a distributed version of an oct-tree); the true owner is defined by which Voronoi cell contains the point in the \emph{current} tessellation, which generally requires explicit geometric containment tests and, in parallel, may involve querying other ranks. Therefore, after each mesh update, we execute a dedicated remapping procedure to replace packets consistently.

We handle these migrations in a dedicated \emph{remapping} phase between the hydrodynamic update and the subsequent Monte Carlo step. Packets always store their physical position in global coordinates, together with a stable cell identifier and a (rank-local) cell index. After rebuilding the tessellation, remapping proceeds in three stages.
\begin{enumerate}
    \item \textbf{Fast index fix-up.} When the cell still exists on the same rank, packets can be updated by matching the stable cell identifier to the new local index. This avoids any time-consuming geometric search in the common case where the tessellation is rebuilt but ownership does not change.
    \item \textbf{Local relabeling.} Packets whose declared cell is no longer valid are re-located using the new geometry. A rank first tries to reassign them locally: if the packet remains inside its declared cell, it is kept unchanged; otherwise, a local spatial index over the mesh points provides a small set of candidate cells that are tested for containment.
    \item \textbf{Inter-rank migration.} Packets that cannot be placed locally are migrated to the ranks to which they belong. We first use a coarse geometric partition to send packets to an approximate owner rank. Remaining unresolved cases are handled by a distributed search: candidate ranks are queried in expanding neighborhoods until one confirms that the packet lies in one of its cells. All exchanges are performed in bulk, and a final validation pass ensures that every packet lies inside its declared cell on its new owner rank.
\end{enumerate}

When load balancing changes the decomposition (i.e., mesh points are exchanged between ranks), packets must be migrated consistently with that exchange. We, therefore, apply the same translation used for mesh-point indices to the packet cell indices and transfer packets to their new owner ranks before the Monte Carlo step begins.

Overall, the moving-mesh remapping ensures that every Monte Carlo step begins from a consistent state in which each packet resides on the rank that owns the Voronoi cell containing its position in the updated tessellation.

\subsubsection{Load Balancing}
The computational cost of the Monte Carlo step is generally distributed differently from that of the hydrodynamic step, because particle traffic concentrates in regions of high optical activity rather than following the mesh-point density. The simulation, therefore, maintains two independent load-balancing decompositions on the same Voronoi mesh: one optimized for the hydrodynamic solver (weighted by cell count or reconstruction cost) and one optimized for the Monte Carlo transport (weighted by radiation workload). Each decomposition assigns a different partitioning of the mesh-generating points across MPI ranks.

Switching between the two decompositions is efficient: because both operate on the same underlying set of mesh-generating points, transitioning from one to the other requires only migrating the points (and their associated hydrodynamic state or photon packets) that change ownership---a partial exchange rather than a full mesh rebuild. The Voronoi tessellation is reconstructed incrementally from the updated point set, reusing the geometric infrastructure already in place. In practice, the fraction of points that migrate between the two decompositions is small when the workload distributions are not radically different, making the switch inexpensive relative to the cost of a full repartitioning.

The user controls how frequently each load balance is reassessed---for example, once every 10 radiation time steps---and supplies a weight function for the Monte Carlo decomposition. In our experience, a simple weight that assigns each cell the number of particle-step events executed inside it during the preceding Monte Carlo step provides a reasonable balance for problems without extreme spatial inhomogeneity.

\subsection{Discussion}\label{sec:discussion}
The per-pair Handler architecture trades memory for concurrency. A na\"ive implementation would allocate a Handler for every pair of ranks, giving $O(P^2)$ Handlers across the communicator. A natural concern is whether this memory overhead is acceptable.

In practice, Handlers are created \emph{lazily}: a Handler between ranks $A$ and $B$ is instantiated only when the two ranks first become communication neighbors and need to exchange particles. On meshes, this usually coincides with geometric adjacency, but the argument is more general: in any spatially local domain decomposition, each rank communicates with only a bounded set of peers. Because such simulations give each rank only $O(1)$ active communication neighbors, only a small fraction of all possible pairs ever materialize. Furthermore, after every load-rebalancing step, the system \emph{shrinks} the buffers of all existing Handlers---reducing those that are no longer between active neighbors to a minimal allocation and reclaiming the freed memory. The dynamic reallocation mechanism (Section~\ref{sec:reallocation}) enlarges buffers on demand when traffic increases. As a result, the aggregate memory footprint at any given time is proportional to the number of \emph{active} communication pairs rather than to $P^2$.

Second, the per-pair design eliminates global locks and enables fully concurrent transfers between disjoint rank pairs. This translates directly into improved scaling behavior: as more ranks are added, each rank's particle workload decreases while the transfer bandwidth remains uncontested. The time saved by avoiding synchronization and lock contention typically far outweighs the additional memory consumed by the Handler metadata.

Third, the Monte Carlo step itself can be memory-intensive, since the number of particle packets in flight at any moment may be large. However, this cost is inherently parallelizable: adding more cores reduces the per-rank particle count proportionally, so memory pressure from the particle population is always relievable by scaling out. The dominant \emph{fixed} memory consumer on each rank is the mesh and hydrodynamic state, not the Monte Carlo communication buffers; even at moderately large rank counts, the Handlers memory footprint is a small fraction of the per-rank memory budget. To give a concrete estimate: each Handler stores a particle ring buffer of $N$ slots at $128$~bytes per slot plus a 16-byte counter pair (\texttt{head} and \texttt{tail}). With a default capacity of $N=1024$ slots, a single Handler occupies approximately $128$\,KB. A rank with ${\sim}20$ active neighbors, therefore, allocates ${\sim}2.6$\,MB for Handlers---a negligible overhead compared with the mesh and hydrodynamic state, which, in our uniform-emission benchmark, amounts to several hundred megabytes per rank.

\subsubsection{Summary of Communication Guarantees}
The combination of the per-pair Handler design, the lock-free SPSC transfer protocol, the reallocation protocol, and the tree-based termination detector provides the following guarantees:
\begin{enumerate}
    \item \textbf{Data-race freedom.} The SPSC ring-buffer invariant structurally prevents data races on the transfer path: only the sender writes \texttt{tail} and particle data, and only the receiver reads particles and writes \texttt{head}. No lock is needed during normal operation; the per-pair Handler architecture ensures that each buffer has exactly one producer and one consumer.
    \item \textbf{Deadlock freedom.} The transfer path is entirely lock-free, eliminating lock-ordering hazards. The reallocation protocol, which falls back to two-sided messaging, breaks potential circular-wait chains through wait-for priority and active progress while blocked (Section~\ref{sec:reallocation}). No rank can be indefinitely stalled waiting for a peer that is itself waiting in a cycle.
    \item \textbf{Starvation freedom.} Time-stamp-ordered service of reallocation requests reduces contention and improves fairness, while the transfer protocol retries after reallocation, so a sender is never permanently blocked by a full buffer.
    \item \textbf{Progress.} The receiver is fully passive during transfers: it issues no matching receive and is never interrupted. A sender can complete a transfer in a bounded number of RDMA operations (one counter read, one or two particle writes, and one atomic increment). Even while blocked on reallocation acknowledgments, a rank continues to serve incoming requests, ensuring system-wide forward progress.
    \item \textbf{Correct termination.} The tree-based counter aggregation and two-phase verification (Section~\ref{sec:termination}) detect global completion without false positives; the verification phase ensures that no in-flight particles are overlooked.
\end{enumerate}

\subsubsection{Limitations of RDMA-exposed Memory}\label{sec:rdma_limitations}
While our scheme supports a high number of particles per rank, as we demonstrate in the benchmarking results (Section~\ref{sec:benchmarking}), it is important to note that RDMA-exposed memory must be \emph{pinned} (locked in physical memory) by the operating system to allow the network adapter to perform direct memory access. This pinning means that RDMA-registered buffers cannot be evicted to swap space and persist until the shrink operation is explicitly invoked or the program terminates, so peak memory usage during a Monte Carlo step may exceed what the simulation strictly requires at any given moment. To mitigate this, we issue the \texttt{madvise} system call with the \texttt{MADV\_DONTNEED} flag after deregistering RDMA-exposed memory, advising the kernel that the region is no longer needed. This call has no effect on pinned pages, which is why it is issued only \emph{after} deregistration.

\section{Numerical Results}\label{sec:benchmarking}
We assess the performance of our low-level RDMA particle-transfer scheme using two complementary benchmarks. The first is a dedicated communication stress test---a uniform-emission benchmark on a three-dimensional unstructured Voronoi mesh. It is designed to isolate the cost of inter-rank particle transfers from physics-related work: every cell in the domain emits photons uniformly into a completely transparent medium with rigid (reflecting) boundaries, so photons free-stream across cell boundaries throughout the entire domain until their time step expires. The second is a cylindrical Hohlraum \textsc{IMC} benchmark, which retains realistic heterogeneous opacities, boundary-driven radiation, temperature evolution, and spatially nonuniform communication traffic. Unless stated otherwise, performance results are obtained with the OFI backend using the \texttt{verbs} provider on InfiniBand hardware; comparisons with the optimized point-to-point backend quantify the benefit of the low-level one-sided implementation over a carefully tuned MPI baseline.

\subsection{Computing Environment}\label{sec:computing}
\textsc{RICH} is written in C++ and was compiled with GCC~15.1.0 (\texttt{g++}). For message passing, we use Open~MPI~4.1.6 (itself built with GCC~15.1.0). The code has no external library dependencies beyond MPI itself, Boost~1.85.0, and, for the native RDMA backends, \texttt{libfabric} (OFI) or \texttt{libibverbs} (IBV). All experiments were performed on the Leonardo supercomputer (CINECA, Italy),\footnote{\url{https://docs.hpc.cineca.it/hpc/leonardo.html}} using the Data Centric General Purpose (DCGP) partition. Each DCGP node contains two Intel Sapphire Rapids Xeon Platinum 8480 processors (56~cores per socket, 112~cores per node). The nodes are interconnected via InfiniBand HDR100 (100\,Gb\,s$^{-1}$) using NVIDIA ConnectX-6 adapters, organized in a Dragonfly+ topology.

\subsection{Uniform-emission Benchmark}\label{sec:uniform_benchmark}

\subsubsection{Setup}\label{sec:uniform_setup}
The benchmark geometry is a three-dimensional cube of side length $L = 10\,\mathrm{cm}$ centered at the origin, discretized with an unstructured Voronoi mesh. The mesh is constructed from $N_{\mathrm{base}}$ randomly sampled points filling the domain. The resulting point cloud is relaxed using five Lloyd iterations and used to build the parallel Voronoi tessellation.

The entire medium is completely transparent: absorption and scattering opacities are identically zero. Photon packets, therefore, free-stream at the speed of light without interaction. The boundary condition is rigid (reflecting): packets that reach any face of the bounding cube are reflected back into the domain, keeping all photons alive until their time step expires. No population control is applied---all surviving packets are carried forward without merging or splitting.

At the start of each cycle, \emph{every} mesh cell injects $n_{\mathrm{emit}}$ photon packets. Each packet is placed at a uniformly random position within its cell and launched with a random isotropic direction at speed~$c$. Because the medium is transparent and the boundaries are reflecting, each photon propagates in straight-line segments across cell boundaries for the full duration of the time step. The number of cell crossings per photon scales with the mesh resolution: finer meshes produce more crossings and, hence, more inter-rank transfers. Uniform emission across the entire domain ensures that every rank generates and receives comparable traffic, providing a spatially homogeneous stress test for the communication layer.

The absence of absorption, scattering, or thermal coupling eliminates all physics-related computational cost. The only meaningful work performed by the code is (i)~propagating photons across cell boundaries, (ii)~transferring particles between ranks when they cross subdomain boundaries, and (iii)~detecting global termination. This design isolates the communication overhead as the dominant cost and provides a clean measurement of the transfer protocol's scalability. Validation on realistic radiation-hydrodynamics problems---including Implicit Monte Carlo transport with full absorption, emission, and scattering physics---is presented in a companion paper (M.~Mizrachi et al. 2026, in preparation).

\subsubsection{Scalability}
Both work-preserving and weak-scaling tests keep the per-core mesh density fixed at $N_{\mathrm{base}} / P = 5000$ cells per core (equivalently 560,000 per node on the 112-core Leonardo nodes), with $n_{\mathrm{emit}} = 5$ photons per cell per cycle and five cycles (the last three of which are timed). Because every cell emits, the total mesh size grows proportionally with the number of cores. The per-photon transport cost is controlled by the time step~$\Delta t$: on a mesh of $N$ total cells, the mean cell spacing scales as $N^{-1/3}$, so each photon's cell crossings per unit time scale as $N^{1/3}$, and the total number of particle-step calls is proportional to $N \cdot n_{\mathrm{emit}} \cdot \Delta t \cdot N^{1/3}$. Work-preserving and weak scaling are obtained by choosing the $\Delta t$ exponent so that total work or per-core work, respectively, remains constant.

\paragraph{Work-preserving scaling.}
A na\"ive strong-scaling design would fix the total number of cells and divide them among an increasing number of cores. However, as the core count grows, the per-core cell count drops proportionally to $1/P$. On our benchmark, such a design is superlinear by a wide enough margin to be useless as a measure of parallel efficiency: doubling the ranks from 2240 to 4480 gives a $2.56\times$ speedup, an apparent efficiency of 128\%, and other configurations behave the same way.\footnote{Fixed mesh of $1.12 \times 10^{7}$ cells, five photons per cell per cycle, $\Delta t = 2 \times 10^{-10}$\,s, OFI backend, timing the last three cycles: 27.75\,s at 2240 ranks against 10.86\,s at 4480. A repeated pair of runs gives $2.45\times$ and a variant emitting 30 photons per cell $2.69\times$.} The gain is not an artifact of the work distribution: the global number of cell crossings is the same in both runs, and the load imbalance, in fact, worsens slightly; yet, the cost of an individual crossing falls from 466\,ns to 352\,ns. What shrinks is the per-rank working set that the transport loop chases through: namely, the local Voronoi mesh consulted for the face-intersection lookups of every crossing together with the buffers holding the packets in flight, from 5000 to 2500 cells and from 107\,MB to 56\,MB on average, while twice as many nodes serve a problem of unchanged size. The reduction in per-crossing cost demonstrates that the conventional strong-scaling experiment changes the single-rank computational efficiency as the local problem becomes smaller. This behavior is consistent with improved memory locality; although, we did not collect hardware counters that would allow us to associate it specifically with cache behavior.

We therefore adopt a different strategy: the per-core cell count is held constant (maintaining a realistic memory footprint and cache effect), while the total transport work is controlled through the time step~$\Delta t$.

Concretely, the total transport work---measured by the aggregate number of particle-step calls across all ranks---is held constant. Because the total mesh size grows linearly with the number of cores~$P$, the time step is scaled as $\Delta t \propto P^{-4/3}$ so that each photon crosses proportionally fewer cells, keeping the global step count fixed at approximately $5.26 \times 10^{10}$ particle-step calls per cycle, or $2.35 \times 10^{7}$ per rank at the 2240-rank baseline. At the baseline of 2240~cores (20~nodes), $\Delta t = 2 \times 10^{-10}$\,s. The logs confirm that this holds across the whole range: at 13,440 ranks, the global crossing count is $5.40 \times 10^{10}$, within 3\% of the 2240-rank value. As more cores are added, the per-core share of transport work decreases proportionally, while the volume of inter-rank transfers grows because additional domain boundaries are introduced by the finer decomposition. Work-preserving scaling, therefore, directly tests the efficiency of the transfer protocol and the overhead of the distributed termination detector under increasing communication pressure, without being confounded by cache effects.

\begin{figure*}
    \centering
    \includegraphics[width=\textwidth]{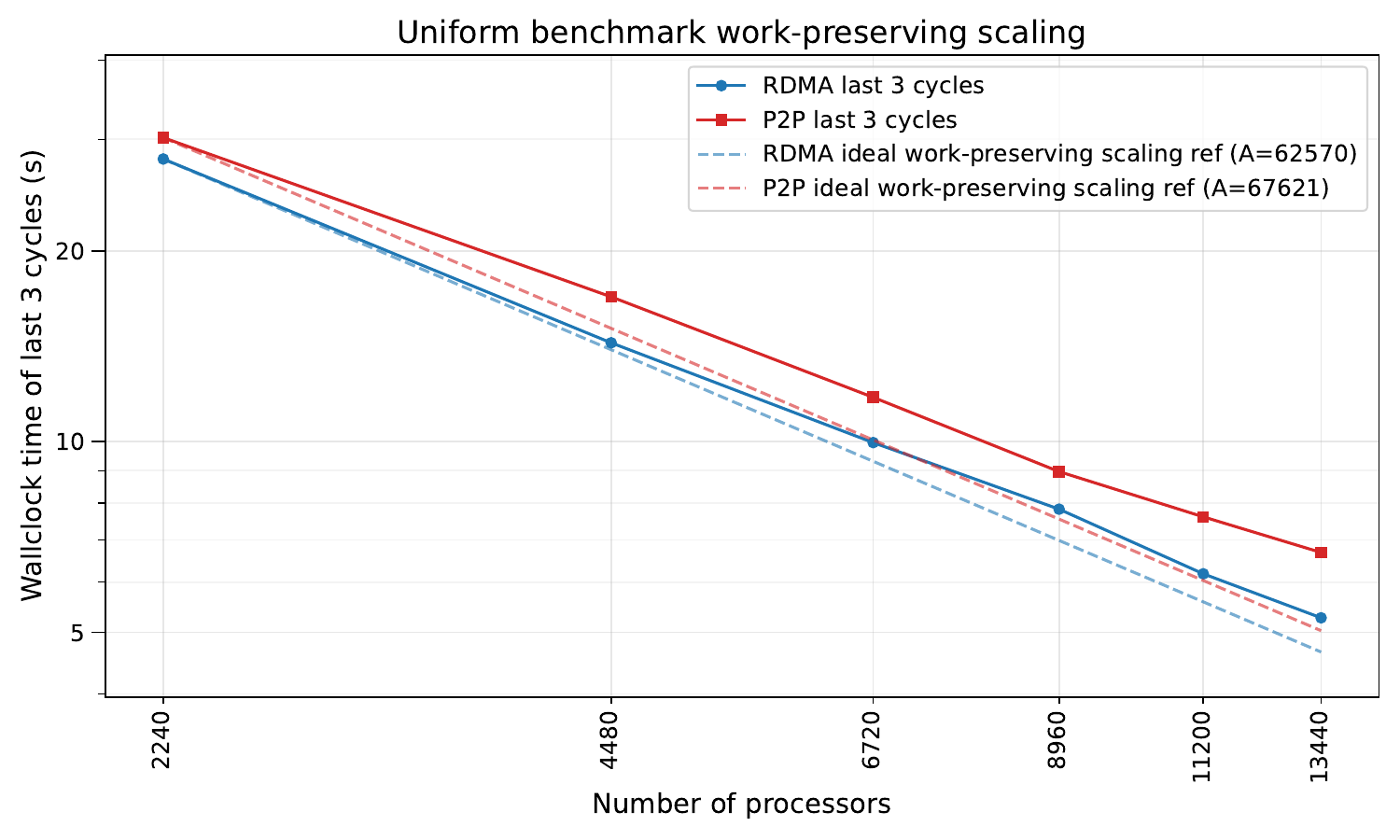}
    \caption{Work-preserving scaling results for the uniform-emission benchmark on a log-log scale. Wall-clock time (average of the last three cycle step times) is shown as a function of the number of MPI ranks for the OFI (RDMA) and point-to-point (P2P) backends. Dashed lines show the ideal $A/P$ scaling anchored at 2240~ranks. The total number of cell crossings is held fixed by scaling the time step as $\Delta t \propto P^{-4/3}$ while keeping 5000 cells per rank.}
    \label{fig:work_preserving_scaling}
\end{figure*}

Figure~\ref{fig:work_preserving_scaling} and Table~\ref{tab:work_preserving_scaling} present the results. The reported wall-clock time is the maximum across all ranks of the per-cycle step time, averaged over the last 3~cycles. Each ideal reference curve has the form $A / P$, where $P$ is the number of processors and $A = T_{\mathrm{ref}} \times P_{\mathrm{ref}}$ is anchored to the measured time at 2240~ranks (separately for each backend).

\begin{table}
\centering
\footnotesize
\renewcommand{\arraystretch}{1.15}
\caption{Work-preserving Scaling Results for the Uniform-emission Benchmark}
\label{tab:work_preserving_scaling}
\setlength{\tabcolsep}{3pt}
\begin{tabular*}{\columnwidth}{@{\extracolsep{\fill}}rrrrrrr@{}}
\hline
 & & \multicolumn{2}{c}{OFI (RDMA)} & \multicolumn{2}{c}{P2P} & RDMA \\
\cline{3-4} \cline{5-6}
Nodes & Ranks & Time & Eff. & Time & Eff. & Speedup \\
 & & (s) & (\%) & (s) & (\%) & \\
\hline
  20 &  2240 & 27.93 & 100.0 & 30.19 & 100.0 & $1.08\times$ \\
  40 &  4480 & 14.32 &  97.5 & 16.93 &  89.2 & $1.18\times$ \\
  60 &  6720 &  9.97 &  93.4 & 11.75 &  85.7 & $1.18\times$ \\
  80 &  8960 &  7.82 &  89.3 &  8.97 &  84.2 & $1.15\times$ \\
 100 & 11,200 &  6.19 &  90.3 &  7.61 &  79.3 & $1.23\times$ \\
 120 & 13,440 &  5.27 &  88.3 &  6.68 &  75.3 & $1.27\times$ \\
\hline
\end{tabular*}
\par\vspace{3pt}
\begin{minipage}{\columnwidth}
\footnotesize
\textbf{Note.} Wall-clock time (s) is the average of the last 3~cycle step times. Efficiency is relative to 2240~ranks (per backend). RDMA speedup is $T_{\mathrm{P2P}} / T_{\mathrm{OFI}}$.
\end{minipage}
\end{table}

Both backends scale well from 2240 to 4480~cores (20-40~nodes), with OFI at 97.5\% efficiency ($1.95\times$ speedup over $2\times$ ideal). At higher core counts, the efficiency decreases gradually: at 11,200~cores (100~nodes) the OFI backend achieves 90.3\% efficiency ($4.52\times$ speedup), while the P2P backend is at 79.3\% ($3.97\times$). At the largest tested scale of 13,440~cores (120~nodes), the OFI backend retains 88.3\% efficiency compared to 75.3\% for P2P. Across the full range, the OFI backend consistently outperforms P2P in both absolute time and scaling efficiency, with the gap widening at larger core counts: the RDMA speedup grows from $1.08\times$ at 2240~ranks to $1.27\times$ at 13,440~ranks.

\paragraph{Weak scaling.}
For the weak-scaling test, the per-core workload is held approximately constant. As described above, the per-core mesh density is the same as for work-preserving scaling (5000 cells per core). The time step is scaled as $\Delta t \propto P^{-1/3}$ so that each photon crosses a similar number of cells per cycle regardless of the total mesh size, keeping the per-core particle-step count approximately constant. At the baseline of 2240~cores, $\Delta t = 2 \times 10^{-10}$\,s.

\begin{figure*}
    \centering
    \includegraphics[width=\textwidth]{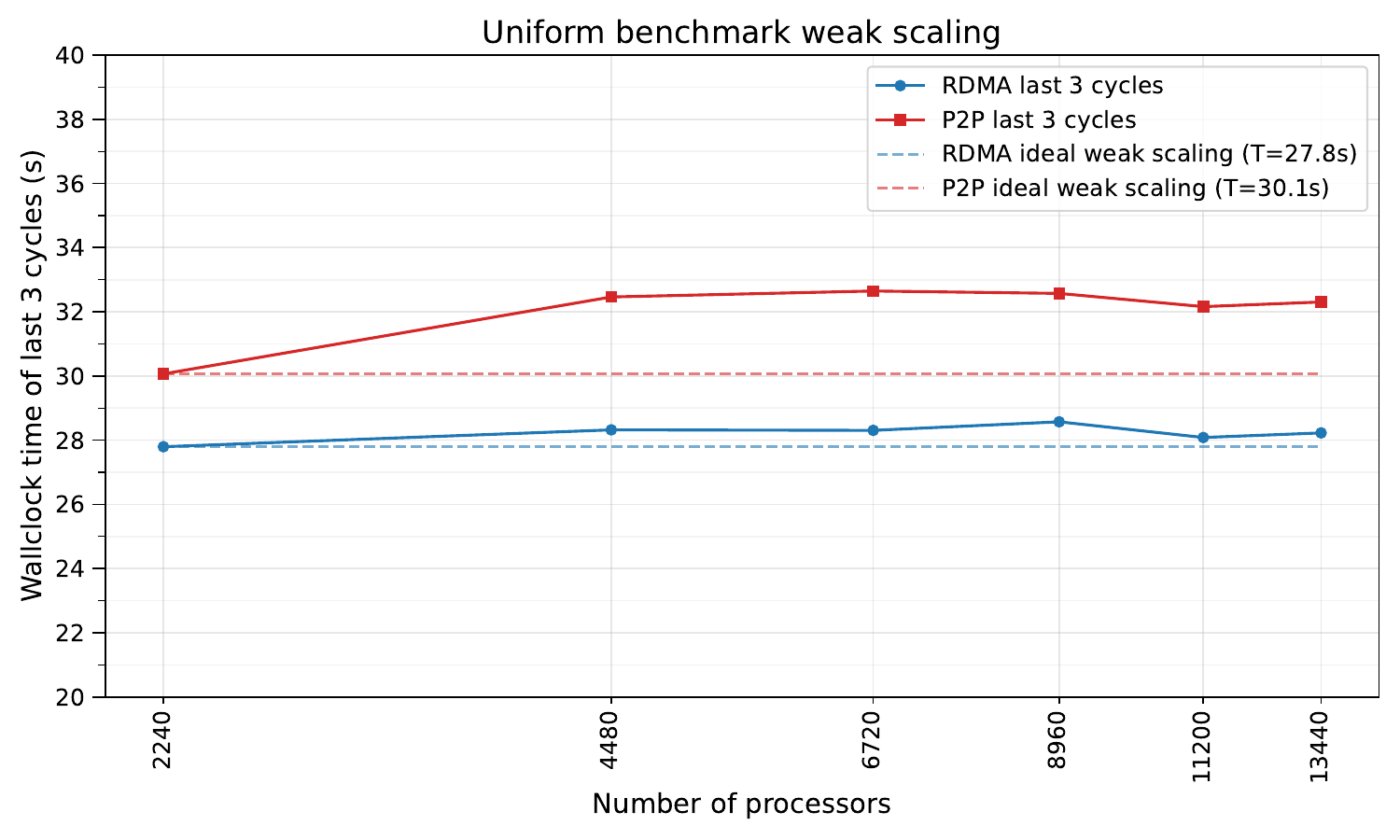}
    \caption{Weak-scaling results for the uniform-emission benchmark. Wall-clock time (average of the last three cycle step times) is shown as a function of the number of MPI ranks, with per-core workload held approximately constant ($N_{\mathrm{base}}/P = 5000$ cells per core, $\Delta t \propto P^{-1/3}$). The dashed line shows the ideal weak-scaling reference (constant time).}
    \label{fig:weak_scaling}
\end{figure*}

Figure~\ref{fig:weak_scaling} and Table~\ref{tab:weak_scaling} present the weak-scaling results. Ideal weak scaling corresponds to a horizontal line: as the problem grows proportionally with the number of processors, the wall-clock time should remain constant. Efficiency is defined as $T_{\mathrm{ref}} / T(P)$, where $T_{\mathrm{ref}}$ is the baseline time at 2240~ranks (20~nodes).

\begin{table}
\centering
\footnotesize
\renewcommand{\arraystretch}{1.15}
\caption{Weak-scaling Results for the Uniform-emission Benchmark}
\label{tab:weak_scaling}
\setlength{\tabcolsep}{3pt}
\begin{tabular*}{\columnwidth}{@{\extracolsep{\fill}}rrrrrrr@{}}
\hline
 & & \multicolumn{2}{c}{OFI (RDMA)} & \multicolumn{2}{c}{P2P} & RDMA \\
\cline{3-4} \cline{5-6}
Nodes & Ranks & Time & Eff. & Time & Eff. & Speedup \\
 & & (s) & (\%) & (s) & (\%) & \\
\hline
  20 &  2240 & 27.80 & 100.0 & 30.07 & 100.0 & $1.08\times$ \\
  40 &  4480 & 28.32 &  98.1 & 32.46 &  92.6 & $1.15\times$ \\
  60 &  6720 & 28.31 &  98.2 & 32.65 &  92.1 & $1.15\times$ \\
  80 &  8960 & 28.57 &  97.3 & 32.57 &  92.3 & $1.14\times$ \\
 100 & 11,200 & 28.08 &  99.0 & 32.16 &  93.5 & $1.15\times$ \\
 120 & 13,440 & 28.23 &  98.5 & 32.31 &  93.1 & $1.14\times$ \\
\hline
\end{tabular*}
\par\vspace{3pt}
\begin{minipage}{\columnwidth}
\footnotesize
\textbf{Note.} Wall-clock time (s) is the average of the last three cycle step times. Efficiency is $T_{\mathrm{ref}} / T(P)$, where $T_{\mathrm{ref}}$ is the baseline time at 20~nodes (2240~ranks) per backend. RDMA speedup is $T_{\mathrm{P2P}} / T_{\mathrm{OFI}}$.
\end{minipage}
\end{table}

The OFI backend demonstrates near-ideal weak scaling across the entire range, with efficiency never falling below 97\%: the wall-clock time varies by $<$3\% from 20 to 120~nodes. The P2P backend exhibits a larger overhead of 7\%--9\% at 40~nodes and above, stabilizing near 92\%--94\% efficiency. Both backends show that the communication layer introduces only modest overhead as the domain is divided more finely. The OFI backend's advantage over P2P is a consistent ${\sim}1.14\times$ speedup across all scales, reflecting the lower per-transfer latency of one-sided RDMA operations.

\subsection{Cylindrical Hohlraum Benchmark}\label{sec:hohlraum_benchmark}
The uniform-emission benchmark intentionally removes most physics so that communication dominates. To complement it, we also use the cylindrical Hohlraum benchmark of \citet{McClarrenUrbatsch2009}, following the setup used in the \textsc{RICH} \textsc{IMC} validation suite (M.~Mizrachi et al. 2026, in preparation). The underlying \textsc{IMC} implementation---including the transport physics, random-walk acceleration, and material coupling---is described in detail in the companion paper. Here, we focus on the communication-layer performance. This benchmark is not a pure communication microbenchmark: it measures the transport layer in a production-like \textsc{IMC} workload with optically thin streaming regions, optically thick absorbing structures, temperature-dependent material response, and strongly nonuniform packet traffic.

\subsubsection{Setup}\label{sec:hohlraum_setup}
The problem is specified in $RZ$ cylindrical geometry and consists of an evacuated cylindrical cavity (the hohlraum) with absorbing walls and internal structures. We implement it as a full three-dimensional simulation by revolving the $RZ$ cross section around the symmetry axis, which we take to be the $x$-axis. The computational domain is
$x \in [0, 1.4]$~cm and $r = \sqrt{y^2 + z^2} \leq 0.65$~cm.

Four material regions define the absorbing structures:
\begin{enumerate}
    \item Left wall: $x \in [0.10, 0.15]$, $r \leq 0.45$~cm.
    \item Capsule: $x \in [0.55, 0.95]$, $r \leq 0.45$~cm.
    \item Right end cap: $x \in [1.35, 1.40]$, $r \leq 0.65$~cm.
    \item Outer cylindrical wall: $x \in [0.10, 1.40]$, $r \in [0.60, 0.65]$~cm.
\end{enumerate}
The absorbing material has opacity
$\kappa_a = 300\, (T/\mathrm{keV})^{-3}$~cm$^{-1}$ and heat capacity
$c_v = 3 \times 10^{15}$~erg\,keV$^{-1}$\,cm$^{-3}$. The vacuum regions have negligible opacity and heat capacity. A blackbody source at $T = 1$~keV drives the left $x$-boundary. The initial temperature is 300~K, and the simulation runs to $t = 10$~ns. No hydrodynamics is applied.

\begin{figure*}[!t]
    \centering
    \begin{tabular}{cc}
        \includegraphics[width=0.48\textwidth]{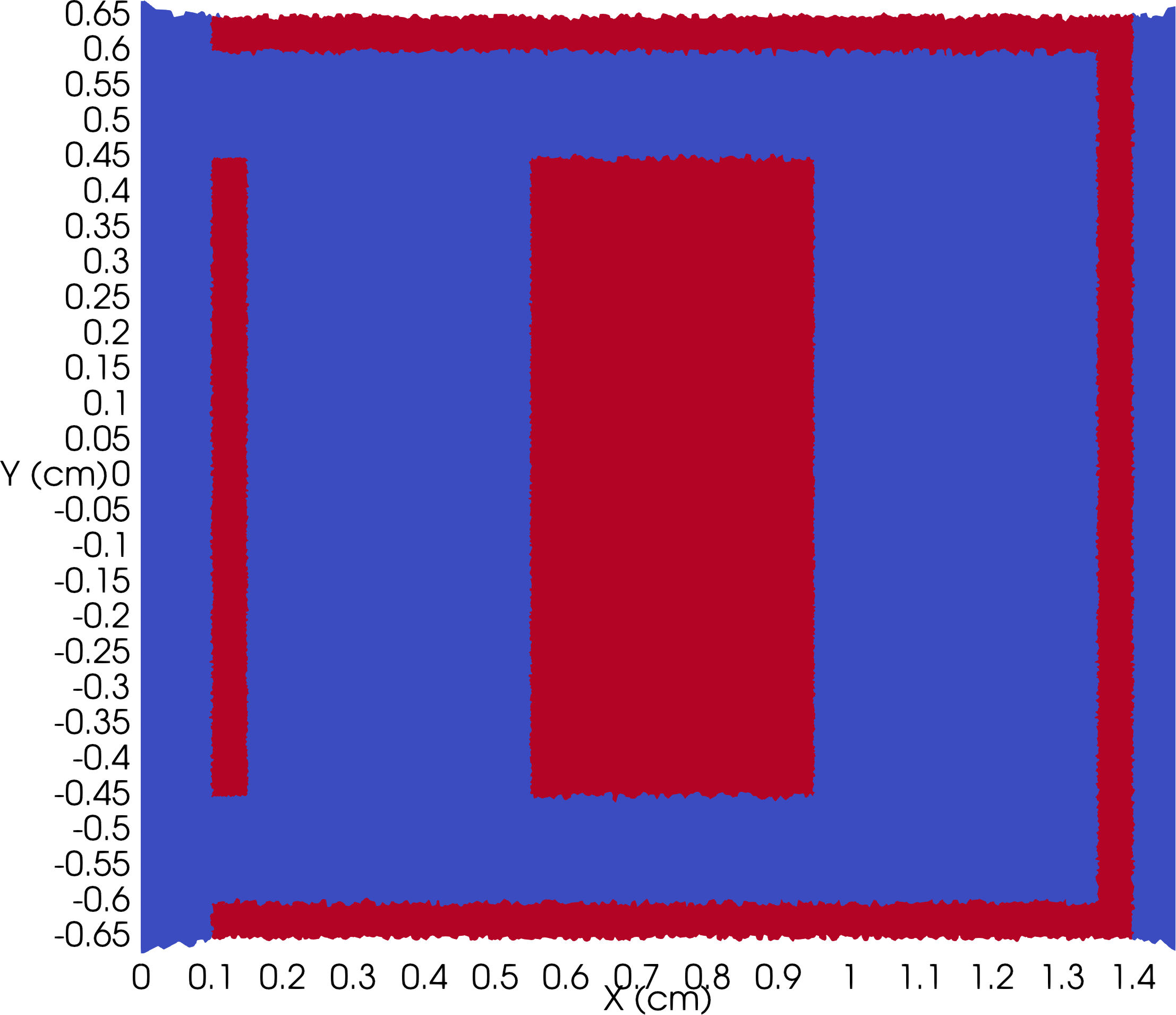} &
        \includegraphics[width=0.48\textwidth]{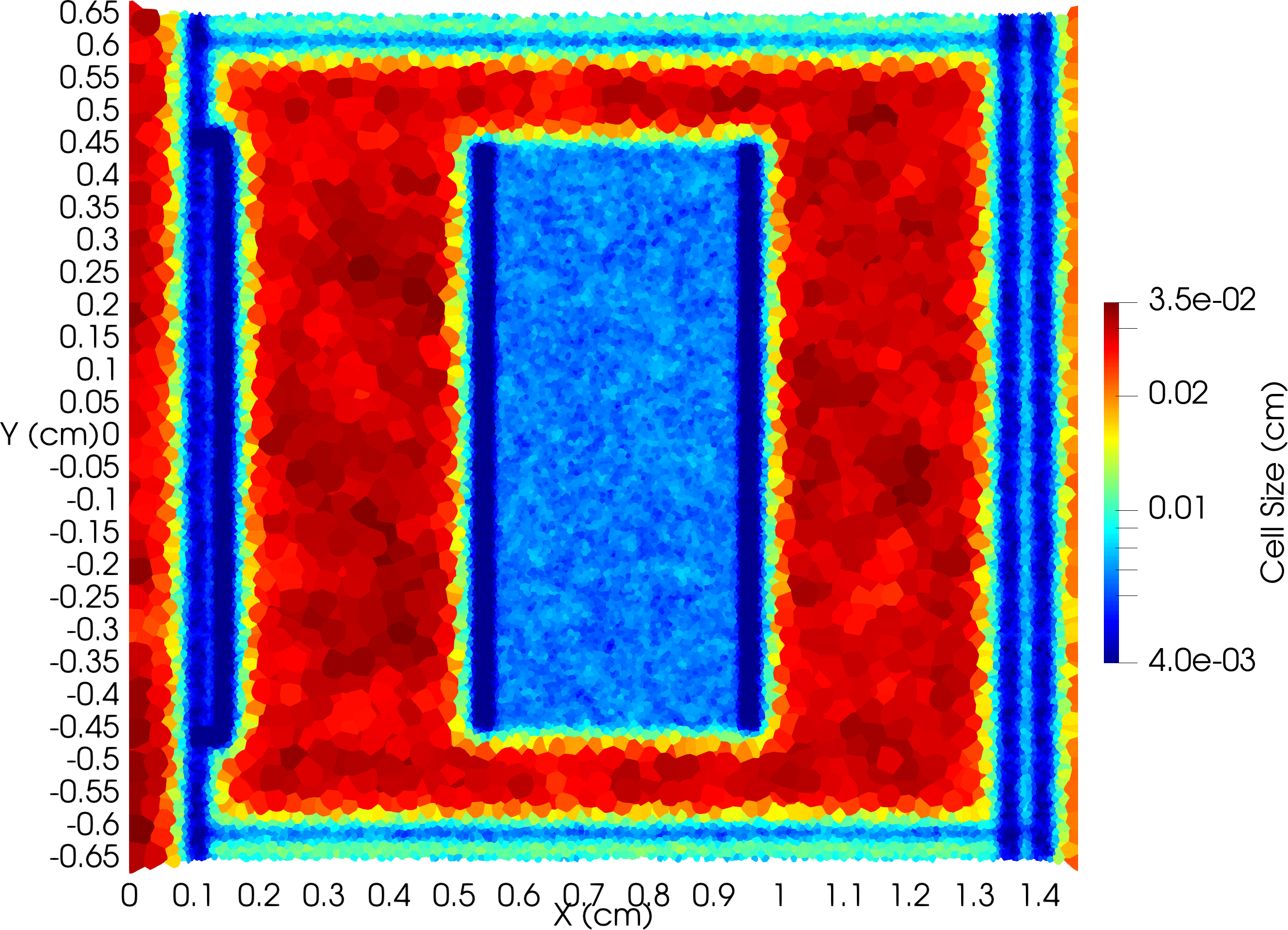} \\
        \includegraphics[width=0.48\textwidth]{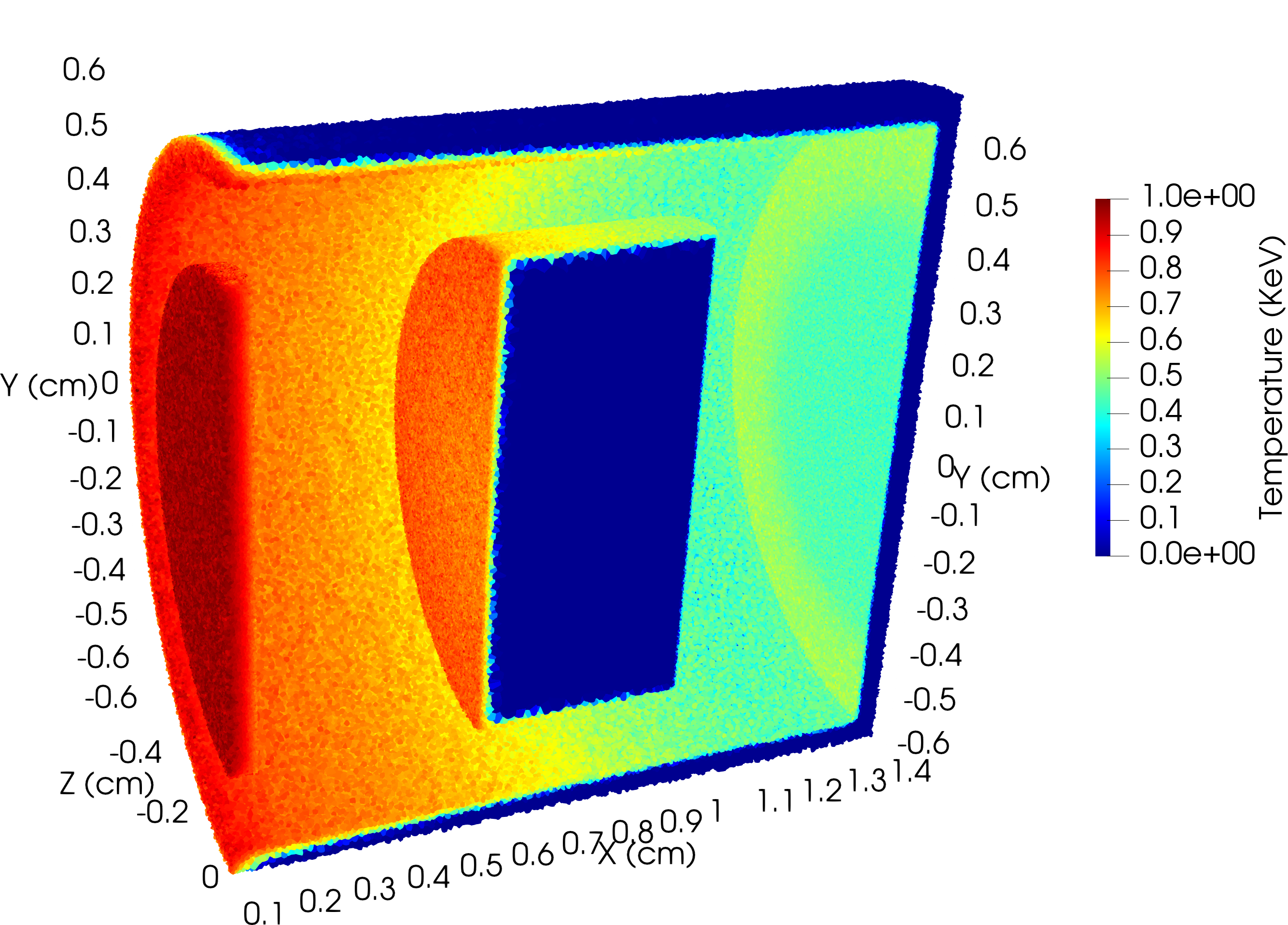} &
        \includegraphics[width=0.48\textwidth]{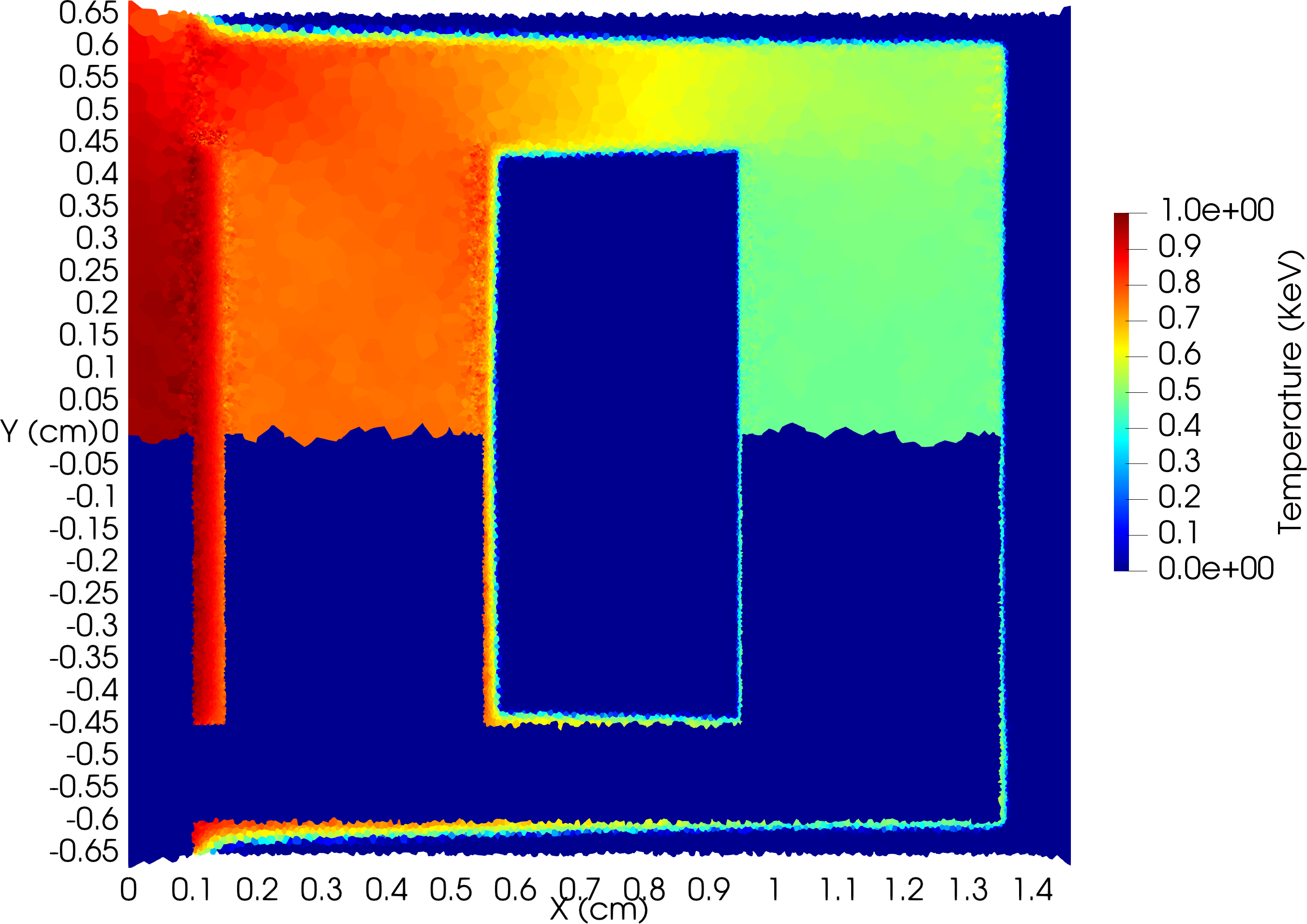}
    \end{tabular}
    \caption{Cylindrical Hohlraum benchmark used as a realistic \textsc{IMC} timing case. Top left: material-type field showing the absorbing wall regions and vacuum channel. Top right: characteristic Voronoi-cell size, with refinement near material-vacuum interfaces. Bottom left: three-dimensional view of the unstructured Voronoi tessellation, showing only the absorbing material regions. Bottom right: representative temperature slices---material temperature (bottom half) and radiation temperature (top half)---illustrating how radiation from the driven $x=0$ boundary heats the capsule and wall regions.}
    \label{fig:hohlraum_benchmark}
\end{figure*}

The Voronoi mesh is generated with refinement near material interfaces: mesh-generating points are concentrated near the boundaries between absorbing and vacuum regions, then relaxed and tessellated. This produces a heterogeneous unstructured mesh of 2,450,000 cells ($N_\mathrm{base} = 50{,}000$) with smaller cells near the sharp opacity transitions and larger cells in the smoother vacuum channel. Figure~\ref{fig:hohlraum_benchmark} demonstrates this mesh structure in $RZ$ geometry, a clipped 3D image of the benchmark, and expected temperature at the end. The mesh is distributed across 4480 MPI ranks, with five new photon packets emitted per cell per time step (up to a maximum of 15). At steady state, each of the 101 Monte Carlo cycles processes approximately 161 million particles: about 100 million census particles carried from the previous time step and roughly 61 million (${\sim}38\%$) newly created via thermal emission. During each cycle, approximately $23.6\%$ of particles are absorbed by the material, about $14.1\%$ escape through the domain boundary, and the remaining ${\sim}62\%$ survive as census particles for the next cycle---maintaining an approximate balance between creation and destruction characteristic of a converged \textsc{IMC} simulation. The \textsc{IMC} random-walk acceleration \citep{Fleck1984} is enabled in optically thick absorbing regions so that the benchmark measures the communication layer in the same accelerated transport mode used for realistic calculations.

\subsubsection{Backend Comparison}\label{sec:hohlraum_timing}
The Hohlraum problem stresses different aspects of the communication layer than the uniform-emission benchmark. Radiation is injected from one side rather than uniformly throughout the domain, so packet density and inter-rank traffic evolve in space and time as the radiation front moves through the vacuum channel and into the absorbing material. The transparent cavity creates long streaming trajectories that cross many cell and rank boundaries, while the optically thick walls and capsule create localized work through repeated material interactions and random-walk events. The resulting workload is both communication-intensive and load-imbalanced, making it a useful end-to-end benchmark for the full particle manager.

\begin{figure*}[!t]
    \centering
    \resizebox{\textwidth}{!}{\begin{tikzpicture}[
  every node/.style={font=\sffamily},
]
  \definecolor{catPhysics}{RGB}{33,102,172}
  \definecolor{catPolling}{RGB}{146,197,222}
  \definecolor{catSendRecv}{RGB}{214,96,77}
  \definecolor{catTerm}{RGB}{255,183,77}
  \definecolor{catRealloc}{RGB}{76,175,80}
  \definecolor{catBusy}{RGB}{255,224,130}

  \def\bh{0.65}
  \def\yP{1.15}   
  \def\yI{0}      

  \fill[catPhysics]  (0,    \yP) rectangle (0.59,  \yP+\bh);
  \fill[catPolling]  (0.59, \yP) rectangle (1.56,  \yP+\bh);
  \fill[catSendRecv] (1.56, \yP) rectangle (3.78,  \yP+\bh);
  \fill[catTerm]     (3.78, \yP) rectangle (4.35,  \yP+\bh);
  \fill[catBusy]     (4.35, \yP) rectangle (5.68,  \yP+\bh);
  \draw[black!50, thin] (0, \yP) rectangle (5.68, \yP+\bh);

  \node[anchor=east, font=\sffamily\footnotesize\bfseries]
    at (-0.12, \yP+\bh/2) {P2P};

  \node[font=\sffamily\tiny, white, rotate=90]
    at (0.295, \yP+\bh/2) {0.59\,s};

  \node[font=\sffamily\tiny, black!70]
    at (1.075, \yP+\bh/2) {0.97\,s};

  \node[font=\sffamily\scriptsize, white]
    at (2.67, \yP+\bh/2) {2.22\,s (39\,\%)};

  \draw[black!50, thin] (4.065, \yP+\bh) -- (4.065, \yP+\bh+0.15);
  \node[above, font=\sffamily\tiny, anchor=south] at (4.065, \yP+\bh+0.12)
    {0.57\,s};

  \node[font=\sffamily\tiny, black!70]
    at (5.015, \yP+\bh/2) {1.33\,s};

  \node[anchor=west, font=\sffamily\small\bfseries]
    at (5.78, \yP+\bh/2) {5.68\,s};

  \draw[decorate, decoration={brace, amplitude=4pt, raise=1pt}, thick, black!60]
    (0.59, \yP+\bh+0.02) -- node[above=5pt, font=\sffamily\tiny\bfseries, black!70]
    {comm.\ 3.76\,s} (4.35, \yP+\bh+0.02);

  \fill[catPhysics]  (0,    \yI) rectangle (0.59,  \yI+\bh);
  \fill[catPolling]  (0.59, \yI) rectangle (0.75,  \yI+\bh);
  \fill[catSendRecv] (0.75, \yI) rectangle (2.67,  \yI+\bh);
  \fill[catTerm]     (2.67, \yI) rectangle (2.98,  \yI+\bh);
  \fill[catRealloc]  (2.98, \yI) rectangle (3.17,  \yI+\bh);
  \fill[catBusy]     (3.17, \yI) rectangle (4.02,  \yI+\bh);
  \draw[black!50, thin] (0, \yI) rectangle (4.02, \yI+\bh);

  \node[anchor=east, font=\sffamily\footnotesize\bfseries]
    at (-0.12, \yI+\bh/2) {OFI (RDMA)};

  \node[font=\sffamily\tiny, white, rotate=90]
    at (0.295, \yI+\bh/2) {0.59\,s};

  \draw[black!50, thin] (0.67, \yI) -- (0.67, \yI-0.15);
  \node[below, font=\sffamily\tiny, anchor=north] at (0.67, \yI-0.15)
    {0.16\,s};

  \node[font=\sffamily\scriptsize, white]
    at (1.71, \yI+\bh/2) {1.92\,s (48\,\%)};

  \draw[black!50, thin] (2.825, \yI) -- (2.825, \yI-0.55);
  \node[below, font=\sffamily\tiny, anchor=north] at (2.825, \yI-0.55)
    {0.31\,s};

  \draw[black!50, thin] (3.075, \yI) -- (3.075, \yI-0.15);
  \node[below, font=\sffamily\tiny, anchor=north] at (3.075, \yI-0.15)
    {0.19\,s};

  \node[font=\sffamily\tiny, black!70]
    at (3.595, \yI+\bh/2) {0.85\,s};

  \node[anchor=west, font=\sffamily\small\bfseries]
    at (4.12, \yI+\bh/2) {4.02\,s};

  \draw[decorate, decoration={brace, amplitude=4pt, raise=1pt, mirror}, thick, black!60]
    (0.59, \yI-0.02) -- node[below=5pt, font=\sffamily\tiny\bfseries, black!70]
    {comm.\ 2.58\,s} (3.17, \yI-0.02);

  \draw[<->, >=stealth, thick, black!60]
    (5.68, \yP) -- node[right, font=\sffamily\scriptsize, xshift=1pt]
    {$1.41\times$} (5.68, \yI+\bh);

  \draw[black!60] (0, -1.20) -- (6.2, -1.20);
  \foreach \x in {0, 1, 2, 3, 4, 5, 6} {
    \draw[black!60] (\x, -1.27) -- (\x, -1.13);
    \node[below, font=\sffamily\scriptsize, black!70] at (\x, -1.27) {\x};
  }

  \node[font=\sffamily\footnotesize, black!70] at (3.0, -1.80)
    {Average per-rank wall-clock time per MC step (s)};

  \pgfmathsetmacro{\ly}{-2.35}
  \def\lsz{0.24}
  \def\lsp{0.10}

  \node[anchor=west, font=\sffamily\tiny\bfseries, black!60] at (0, \ly+\lsz/2) {Computation:};
  \fill[catPhysics]  (1.50, \ly) rectangle (1.50+\lsz, \ly+\lsz);
  \node[anchor=west, font=\sffamily\tiny]
    at (1.50+\lsz+\lsp, \ly+\lsz/2) {Physics};

  \node[anchor=west, font=\sffamily\tiny\bfseries, black!60] at (3.20, \ly+\lsz/2) {Busy waiting:};
  \fill[catBusy]  (4.60, \ly) rectangle (4.60+\lsz, \ly+\lsz);
  \node[anchor=west, font=\sffamily\tiny]
    at (4.60+\lsz+\lsp, \ly+\lsz/2) {Imbalance (Loop Overhead)};

  \pgfmathsetmacro{\lyy}{\ly - 0.42}
  \node[anchor=west, font=\sffamily\tiny\bfseries, black!60] at (0, \lyy+\lsz/2) {Communication:};
  \fill[catPolling]  (1.50, \lyy) rectangle (1.50+\lsz, \lyy+\lsz);
  \node[anchor=west, font=\sffamily\tiny]
    at (1.50+\lsz+\lsp, \lyy+\lsz/2) {MPI Progress};
  \fill[catSendRecv] (3.20, \lyy) rectangle (3.20+\lsz, \lyy+\lsz);
  \node[anchor=west, font=\sffamily\tiny]
    at (3.20+\lsz+\lsp, \lyy+\lsz/2) {Send/Recv};
  \fill[catTerm]     (4.60, \lyy) rectangle (4.60+\lsz, \lyy+\lsz);
  \node[anchor=west, font=\sffamily\tiny]
    at (4.60+\lsz+\lsp, \lyy+\lsz/2) {Termin.};
  \fill[catRealloc]  (6.00, \lyy) rectangle (6.00+\lsz, \lyy+\lsz);
  \node[anchor=west, font=\sffamily\tiny]
    at (6.00+\lsz+\lsp, \lyy+\lsz/2) {Realloc.};
\end{tikzpicture}}
    \caption{Average per-rank wall-clock time breakdown for a representative Monte Carlo step of the cylindrical Hohlraum \textsc{IMC} benchmark on 40~nodes (4480~ranks), comparing the OFI (RDMA) and P2P backends. All values are averages across ranks. \textbf{Computation:} \emph{Physics}---the \texttt{physics->step()} call (face-intersection geometry, opacity lookup, scattering, and energy deposition). \textbf{Communication:} \emph{MPI Progress}---per-iteration overhead of MPI progress calls, cell-move bookkeeping, particle-list management, and send-buffer queuing within the inner particle loop; \emph{Send/Recv}---send and receive calls, buffer management, and flushing of aggregated particle buffers to the network; \emph{Termination}---probing and advancing the tree-based distributed termination counter; \emph{Realloc.}---asynchronous buffer reallocation progress (one-sided backends only). \textbf{Busy waiting:} accumulated per-iteration overhead of timing instrumentation and loop control across millions of main-loop iterations. Both backends perform the same $9.31 \times 10^{9}$ particle steps; the physics time is identical. The $1.41\times$ speedup originates from reduced MPI progress overhead and lower communication costs.}
    \label{fig:hohlraum_breakdown}
\end{figure*}
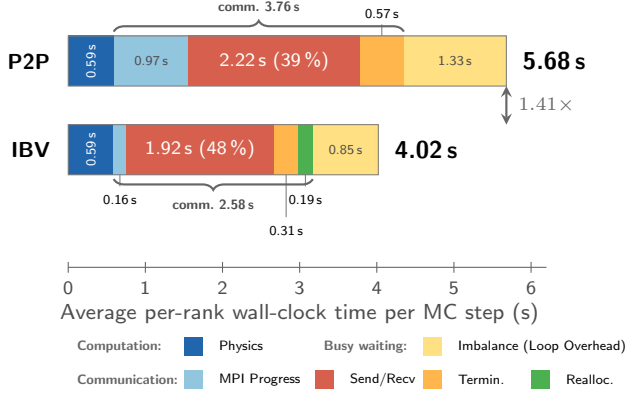

To compare the OFI and P2P backends under realistic \textsc{IMC} traffic, we run the same Hohlraum problem with both backends on 40~nodes (4480~ranks). Figure~\ref{fig:hohlraum_breakdown} decomposes the \emph{average} per-rank wall-clock time for a representative Monte Carlo step into six categories (see the caption for definitions). All values are averages across ranks for a single representative stabilized time step.

The OFI backend completes each Monte Carlo step in 4.02\,s versus 5.68\,s for P2P, a $1.41\times$ speedup. Both backends execute the same \textsc{IMC} physics code and perform the same total number of particle steps ($9.31 \times 10^{9}$), so the \emph{physics} time---face-intersection calculations, opacity lookups, scattering events, and energy deposition---is identical at 0.59\,s per rank. The largest single category in both backends is \emph{send/recv}: the combined cost of scanning per-rank buffers for incoming particles and flushing aggregated send buffers to the network. This accounts for 1.92\,s (48\%) for OFI and 2.22\,s (39\%) for P2P. Because the particle workload is unevenly distributed---the most loaded rank processes ${\sim}5\times$ more particles than the average---much of the receive-side time is effectively idle waiting due to load imbalance. The category showing the largest relative difference between backends is \emph{MPI progress}---the per-iteration overhead of MPI progress calls, cell-move bookkeeping, particle-list management, and send-buffer queuing within the inner particle loop---which costs 0.16\,s in OFI versus 0.97\,s in P2P ($6.1\times$). This gap arises because the P2P backend must explicitly drive MPI receive progress, resulting in ${\sim}4.4\times$ more main-loop iterations and proportionally more per-iteration overhead. The remaining costs---\emph{termination} (the distributed completion counter, ${\sim}$8\%--10\%), \emph{reallocation} (one-sided only, 4.7\%), and \emph{busy waiting} (${\sim}$21\%--24\%)---are individually modest but consistently lower in the OFI backend.

In summary, the Hohlraum benchmark demonstrates that the OFI backend achieves a $1.41\times$ wall-clock speedup over the P2P backend at 4480~ranks, with lower overhead in every nonphysics category (Figure~\ref{fig:hohlraum_breakdown}). The \textsc{IMC} implementation that underpins these benchmarks---including the transport physics, random-walk acceleration, and coupling to the Voronoi mesh---is described in detail in the companion paper (M.~Mizrachi et al. 2026, in preparation).

\section{Conclusion}\label{sec:conclusion}
We have presented \textsc{STORM}, a publicly available, open-source library providing a scalable, mesh-independent communication layer for Monte Carlo particle transport in domain-decomposed, distributed-memory simulations. The core design---a per-pair SPSC ring-buffer Handler architecture with lock-free one-sided RDMA transfers, local aggregation, deadlock-free buffer reallocation, and tree-based termination detection---eliminates cross-sender contention, removes all locking from the common transfer path, avoids deadlock by construction, and leaves the receiving rank entirely passive during transfers.

\textsc{STORM} is independent of the spatial discretization and transport physics: it requires only that the application determine when a particle leaves the local subdomain and identify the destination rank. Four interchangeable backends---OFI (\texttt{libfabric}), native InfiniBand Verbs, MPI~RMA, and optimized point-to-point messaging---accommodate a wide range of cluster hardware without modifying application code.

We validated \textsc{STORM} within the open-source \textsc{RICH} code using a uniform-emission benchmark---a deliberately adversarial stress test with a transparent medium and reflecting boundaries, designed to maximize inter-rank particle traffic and isolate the communication cost from physics-related work. Even under these conditions, and on a particularly demanding hardware configuration (112~cores sharing a single 100\,Gb/s NIC per node), the OFI backend maintains 88\% work-preserving scaling parallel efficiency at 13,440~cores (120~nodes) and near-ideal weak-scaling efficiency above 97\% across the entire range from 20--120~nodes. To our knowledge, this combination of scaling efficiency under adversarial communication load at this core count is unique among Monte Carlo transport implementations.

It is important to note that the uniform-emission benchmark represents an adversarial scenario for communication scalability: the medium is entirely transparent, and the boundaries are reflecting, so every photon free-streams across the domain without scattering or absorption for the full duration of the time step. In realistic radiation-transport problems, photons undergo interactions that keep them local for longer periods, substantially reducing the volume of inter-rank transfers relative to useful physics work. A per-step time breakdown on a cylindrical Hohlraum \textsc{IMC} benchmark---with heterogeneous opacities, boundary-driven radiation, and spatially nonuniform traffic---confirms this: the physics computation (face intersection, scattering, energy deposition) is identical between backends and accounts for ${\sim}15\%$ of the average rank's time, while the dominant cost is receive polling, much of which reflects idle waiting due to load imbalance rather than active communication. The $1.41\times$ OFI speedup at 4480~ranks originates from reduced MPI progress overhead ($6.1\times$ lower), more efficient receive polling, and lower communication costs---advantages of the passive RDMA receive model over explicit MPI progress calls. Detailed physics validation for the Hohlraum problem is presented in M.~Mizrachi et al. (2026, in preparation).

A further contributing factor to the observed overhead is network contention: each node's single 100\,Gb\,s$^{-1}$ InfiniBand NIC is shared among all 112~cores, providing less than 1\,Gb\,s$^{-1}$ of effective per-core bandwidth under uniform traffic. This is a particularly demanding configuration; on systems with fewer cores per node, or equipped with higher-bandwidth network adapters (e.g., HDR~200\,Gb\,s$^{-1}$ or NDR~400\,Gb\,s$^{-1}$), the per-core bandwidth would increase substantially, and we expect the communication overhead to decrease correspondingly, yielding significantly higher scaling efficiency than reported here.

Notably, both the OFI and P2P backends confirm that the Handler architecture and lock-free protocol maintain their scalability regardless of the underlying transport mechanism; although, the OFI backend consistently achieves higher efficiency at larger core counts. This should be interpreted as a comparison between an application-specific low-level RDMA protocol and generic MPI point-to-point semantics, not as a comparison between RDMA hardware and an MPI path that lacks RDMA. The MPI implementation may use RDMA internally, but it still exposes matched-message progress, posted receives, and library-controlled completion. The OFI backend's primary benefit is the direct control it gives over remote-buffer layout, memory registration, operation ordering, and completion granularity, yielding a consistent speedup in absolute wall-clock time---ranging from $1.14\times$--$1.27\times$ over the optimized P2P backend on the uniform-emission benchmark at 40~nodes and above, and $1.41\times$ on the realistic Hohlraum \textsc{IMC} workload at 40~nodes.

The OFI backend's reliance on \texttt{libfabric} rather than a hardware-specific API ensures broad portability: the same code path supports InfiniBand (via the \texttt{verbs} provider), Cray Slingshot (\texttt{cxi}), AWS Elastic Fabric Adapter (\texttt{efa}), and Ethernet (\texttt{tcp/rxm}). A native IBV backend is also available for direct InfiniBand Verbs access where desired, and the MPI RMA and point-to-point backends serve as universal fallbacks on platforms without \texttt{libfabric}.

Looking ahead, \textsc{STORM} removes a key barrier to scaling Monte Carlo transport in astrophysical multiphysics codes. High-fidelity simulations of supernovae, neutron-star mergers, and accretion disks increasingly demand coupled radiation-hydrodynamics with energy- and angle-resolved photon or neutrino transport on dynamically evolving meshes. By decoupling the communication substrate from both the physics model and the mesh representation, \textsc{STORM} can serve as a reusable building block for such applications, allowing developers to scale to larger core counts without redesigning the particle-transfer infrastructure.

\begin{acknowledgments}
This work used computational resources awarded by the EuroHPC Joint Undertaking (project EHPC-DEV-2026D04-204), on Leonardo hosted by CINECA.
\end{acknowledgments}

\section*{Code Availability}
\textsc{STORM} is publicly available as open-source software under the BSD 3-Clause licence. It is distributed both as a standalone package and as an integrated module within the \textsc{RICH} three-dimensional moving-mesh hydrodynamics code at \url{https://gitlab.com/eladtan/RICH}. \textsc{STORM} interfaces with the \textsc{MadVoro} parallel Voronoi tessellation builder \citep{Mizrachi2025madvoro} for applications on unstructured moving meshes, but it can be coupled to any mesh representation. The version associated with this paper is permanently archived on Zenodo \citep{Mizrachi_STORM_2026}. Ongoing development is available at \url{https://github.com/maormizrachi/STORM}. The code may be freely used and modified; we request that any published work that uses \textsc{STORM} cite this paper.

\bibliography{example}{}
\bibliographystyle{aasjournalv7}

\end{document}